\documentclass[11pt,a4paper]{article}
\usepackage[utf8]{inputenc}
\usepackage{graphicx}
\usepackage{amsmath,amssymb,mathtools}
\usepackage{hyperref}
\hypersetup{colorlinks=true,linkcolor=blue,citecolor=blue,urlcolor=blue}
\usepackage{geometry}
\usepackage{multirow}
\date{}

\begin{document}

\title{{\bf Chaos and epoch structure in the deformed Mixmaster universe}}

\author{Babak Vakili\thanks{email:
ba.vakili@iau.ac.ir}\\\\{\small {\it
Department of Physics, CT.C., Islamic Azad
University, Tehran, Iran}}} \maketitle

\begin{abstract}
We study the dynamics of the Bianchi~IX (Mixmaster) universe under classical
polymerization and generalized uncertainty principle (GUP) deformation
of the Poisson brackets. Starting from the Misner Hamiltonian, we derive the
effective equations of motion with both modifications and analyze the duration
of Kasner epochs as a probe of dynamical behavior. Our results show that GUP
corrections typically shorten the epochs, leading to more frequent wall
collisions, whereas polymer corrections prolong them and suppress successive
bounces. At leading order, the combined deformation produces an additive shift
that interpolates between these two trends. While the billiard picture remains
robust, the strength of Mixmaster chaos becomes sensitive to the deformation
parameters. These results illustrate how Planck-scale corrections may either
enhance or suppress cosmological chaos, offering a controlled framework for
exploring early-universe dynamics.
\vspace{5mm}\noindent\vspace{0.8mm}\newline Keywords: Bianchi IX cosmology; Mixmaster chaos; Polymer quantization; 
GUP; Deformed Poisson brackets; Kasner epochs
\end{abstract}

\section{Introduction}

The study of anisotropic cosmological models has played a central role in
understanding the possible behavior of the early universe beyond the idealized
homogeneous and isotropic Friedmann–Robertson–Walker (FRW) geometries. Among
them, the Bianchi~IX model --- also known as the Mixmaster universe --- has
attracted particular attention due to its highly nontrivial dynamics near the
cosmological singularity. In the seminal works of
Belinski, Khalatnikov and Lifshitz (BKL)~\cite{belinskii1970,belinskii1982},
the asymptotic approach to the singularity in Bianchi~IX was shown to be
governed by an infinite sequence of Kasner epochs interspersed with chaotic
bounces against exponential potential walls. This chaotic behavior, often
described through the ``cosmological billiard'' picture~\cite{damour2003billiards}, has
established the Mixmaster model as a prototype for studying deterministic chaos
in general relativity.

Subsequent analyses confirmed the chaotic nature of the Bianchi~IX dynamics
using both Hamiltonian and dynamical systems techniques
(e.g.~\cite{misner1969,barrow1982,ruelle1983,berger1993}), with the BKL map
providing a simple discrete description of the evolution of Kasner exponents.
Beyond its mathematical richness, this model serves as a theoretical laboratory
to test how fundamental corrections may alter the behavior of spacetime near
the Planck regime. In particular, questions such as whether chaos persists, is
suppressed, or is modified by quantum effects remain of great interest in
quantum cosmology.

Two prominent classes of Planck-scale inspired modifications have been widely
studied in simpler cosmological models: (i) polymer quantization, motivated by
loop quantum gravity, which discretizes phase space variables and often leads
to the replacement $p \mapsto \frac{1}{\mu}\sin(\mu p)$ in Hamiltonians, and
(ii) generalized uncertainty principles (GUP), which introduce deformed Poisson
brackets of the form $\{q,p\} = 1 + \beta p^2$, capturing features expected
from various quantum gravity scenarios
\cite{ashtekar2006,bojowald2007,maggiore1993,kempf1995,scardigli1999}.
Each of these frameworks has been extensively analyzed in isotropic and Bianchi~I
settings, where they modify the approach to singularities, affect the effective
Friedmann equations, and sometimes resolve or soften classical divergences
\cite{chiou2007,husain2004,barbero2010,hossenfelder2013}.

Much less attention, however, has been given to the interplay between these
two deformations in more complex anisotropic models such as Bianchi~IX. Given
the Mixmaster's intrinsically chaotic structure, it provides an ideal testing
ground to explore whether GUP and polymer effects reinforce, compete with, or
neutralize one another in shaping early-universe dynamics. In particular, the
duration of Kasner epochs and the stochasticity of the BKL map constitute
natural probes to assess how the characteristic chaotic features are altered.

The aim of the present work is therefore to perform a systematic analysis of
the Bianchi~IX dynamics under combined polymer and GUP-type deformations. After
reviewing the classical Mixmaster Hamiltonian and its well-known chaotic
properties, we introduce the classical polymer prescription and deformed
Poisson brackets, derive the effective equations of motion, and compute the
leading-order corrections to the epoch durations. These analytic results are
then used to assess qualitatively how chaos is modified, and are supported by
sample numerical evaluations. We find that GUP corrections tend to shorten the
epochs and increase bounce frequency, whereas polymer corrections prolong the
epochs and suppress bounces; in the combined case the effects add at first
order, leading to a competition between the two trends.

\section{Review of the classical Mixmaster model}
The Bianchi~IX cosmological model, often referred to as the Mixmaster Universe \cite{misner1969}, represents the most general homogeneous but anisotropic solution of Einstein's field equations with a closed spatial topology $S^3$. Simpler models, such as Bianchi I (flat) and Bianchi II (single-wall potential), appear as limiting cases of the full Bianchi IX dynamics. This hierarchical relation highlights the role of the Mixmaster model as a ``laboratory'' for studying the onset of chaos in gravitational systems. Its dynamics capture the essential features of the approach to the cosmological singularity, exhibiting a complex sequence of Kasner epochs and chaotic transitions. It provides a paradigmatic example of chaotic dynamics in general relativity, and plays a central role in the BKL description of generic spacelike singularities \cite{belinskii1970}. A clear understanding of the classical Mixmaster dynamics is crucial for any attempt to quantize or modify the model. In particular, the billiard picture and the structure of the Hamiltonian potential will provide the natural reference against which polymer quantization or other quantum-gravity-inspired deformations can be compared.

The spacetime metric can be expressed in Misner's parametrization as \cite{misner1969quantum}

\begin{equation}
ds^2 = -N^2(t)\, dt^2 + e^{2\alpha(t)} \, e^{2\beta_{ij}(t)} \, \omega^i \omega^j,
\end{equation}
where $N(t)$ is the lapse function, $\alpha(t)$ encodes the isotropic volume degree of freedom (related to the spatial volume \(V \propto e^{3\alpha}\)), 
$\beta_{ij}$ is a traceless diagonal matrix describing anisotropies and $\omega^i$ 
are the invariant one-forms on $S^3$, satisfying the \(SU(2)\) Lie algebra

\begin{equation}
d\omega^i = \frac{1}{2} \epsilon^{i}{}_{jk} \, \omega^j \wedge \omega^k .
\end{equation}
It is convenient to introduce the Misner variables

\begin{equation}
\beta_{ij} = \mathrm{diag} \left( \beta_+ + \sqrt{3}\beta_-,\ \beta_+ - \sqrt{3}\beta_-,\ -2\beta_+ \right) ,
\end{equation}
where $(\beta_+,\beta_-)$ measuring the two independent anisotropy modes.

In terms of $(\alpha, \beta_+, \beta_-)$ and their canonical momenta $(p_\alpha, p_+, p_-)$, 
the Hamiltonian constraint for the vacuum Bianchi~IX model reads \cite{ryan1972homogeneous}

\begin{equation}\label{Hamil}
\mathcal{H} = -p_\alpha^2 + p_+^2 + p_-^2 + e^{4\alpha} V(\beta_+, \beta_-) \approx 0,
\end{equation}
where the potential term is
\begin{equation}
V(\beta_+, \beta_-) = \frac{1}{3}\left[ 
e^{-8\beta_+} - 4 e^{-2\beta_+} \cosh(2\sqrt{3}\beta_-) 
+ 2 e^{4\beta_+} \left( \cosh(4\sqrt{3}\beta_-) - 1 \right)
\right].
\end{equation}
The potential $V(\beta_+,\beta_-)$ consists of three steep exponential walls 
that enclose a triangular domain in the $(\beta_+,\beta_-)$ plane. 
In the limit $\alpha \to -\infty$ (approaching the singularity), 
the walls become effectively infinitely steep and the dynamics of the 
universe point in the $(\beta_+,\beta_-)$ space can be approximated by 
a free geodesic motion interrupted by specular reflections off the walls.

In suitable Misner--Chitr\'e coordinates, this motion corresponds to 
a billiard on a two-dimensional Lobachevsky (hyperbolic) space \cite{damour2003billiards}. 
The triangular billiard is of finite volume and negatively curved, 
which is a well-known setting for strong chaos. 
The Mixmaster dynamics in this approximation exhibits: (i) Sensitive dependence on initial conditions, quantified by positive Lyapunov exponents, (ii) Ergodicity within the allowed billiard domain and (iii) Stochastic Kasner transitions, where the sequence of $u_n$ values in the BKL map behaves like a random process with a well-defined probability distribution.

The billiard picture not only provides an intuitive geometric understanding of the 
chaotic behavior but also facilitates analytical estimates for quantities such as the 
Lyapunov exponent and the statistical distribution of Kasner epochs.
This feature will be particularly relevant for our later analysis, 
where the polymer and deformed-bracket modifications are expected to alter 
the effective wall geometry and thereby influence the chaotic properties.

\begin{figure}
\includegraphics[width=2.5in]{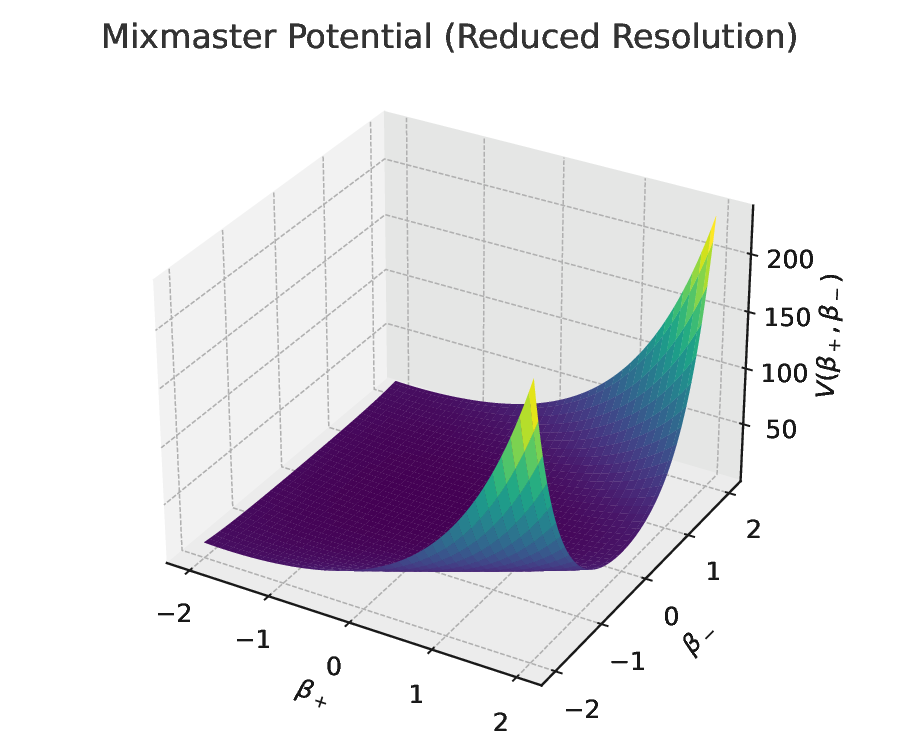}\hspace{25mm}\includegraphics[width=2in]{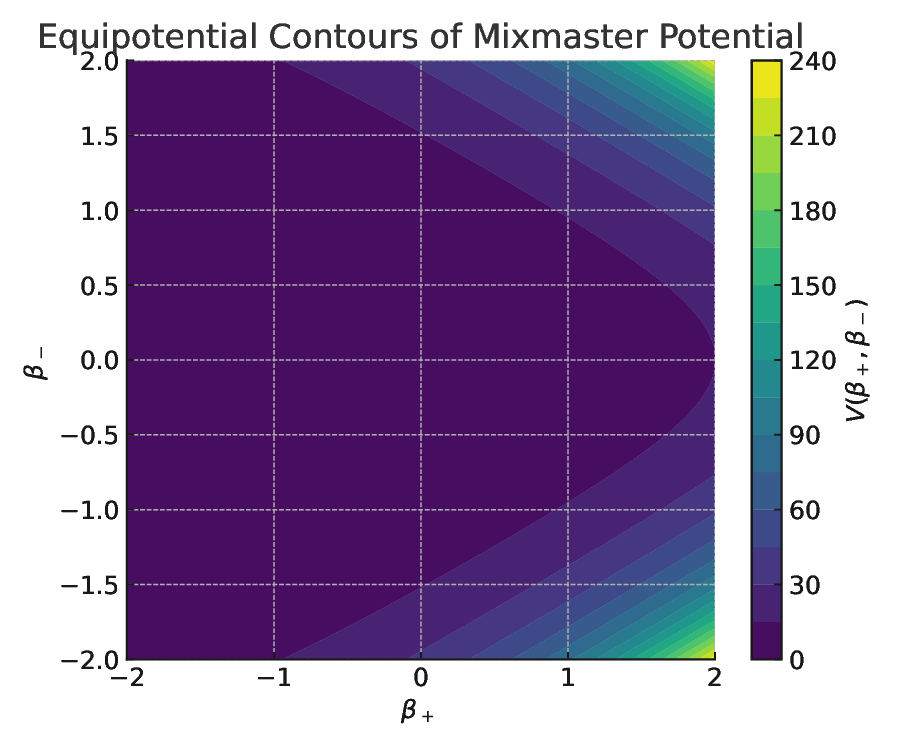}\\
\caption{Left: Three-dimensional representation of the Mixmaster (Bianchi IX) potential $V(\beta_{+},\beta_{-})$ as a function of the Misner anisotropy parameters, showing the steep triangular walls that confine the dynamics in the anisotropy plane. Right: Contour plot of the same potential, where the triangular billiard region becomes apparent and is commonly used to analyze the chaotic geodesic motion of the universe point in minisuperspace.}\label{fig1}
\end{figure}

Figure~\ref{fig1} illustrates the structure of the Bianchi IX (Mixmaster) potential $V(\beta_{+},\beta_{-})$ in the anisotropy plane. 
The left panel presents a full three-dimensional view, where the triangular symmetry of the potential walls is clearly visible. These walls act as impenetrable barriers for the universe point, confining its dynamics to a compact region of minisuperspace. 
The right panel shows the corresponding contour plot, in which the triangular billiard domain becomes evident. 
This billiard representation is particularly useful for analyzing the chaotic properties of the Mixmaster universe, since the dynamics near the singularity can be approximated as a sequence of free geodesic motions interrupted by specular reflections from the potential walls — a behavior closely related to the properties of chaotic billiards in mathematical physics \cite{barrow1982, chernoff1983mixmaster, cornish1997mixmaster}. Indeed, what happened is that near the singularity, the potential walls in the $(\beta_+,\beta_-)$ plane become infinitely steep, so that the dynamics can be approximated by a free motion interrupted by specular reflections against the walls. This leads to the celebrated ``cosmological billiard'' picture, in which the trajectory of the system resembles that of a particle moving inside a triangular billiard table in a hyperbolic space. The sequence of reflections encodes a chaotic map, which is the source of the highly irregular behavior of the Mixmaster model.

When the universe point in minisuperspace is far from the potential walls, the dynamics reduces to the Kasner solution
\begin{equation}
ds^2 \approx -dt^2 + t^{2p_1} dx^2 + t^{2p_2} dy^2 + t^{2p_3} dz^2,
\end{equation}
with the Kasner exponents $(p_1,p_2,p_3)$ satisfying
\begin{equation}
p_1 + p_2 + p_3 = 1, \qquad p_1^2 + p_2^2 + p_3^2 = 1.
\end{equation}
The approach to the singularity involves a sequence of Kasner epochs, 
interrupted by bounces off the potential walls. The change of Kasner parameters 
is governed by the BKL map

\begin{equation}
u_{n+1} = 
\begin{cases}
u_n - 1, & u_n \ge 2, \\
\frac{1}{u_n - 1}, & 1 < u_n < 2,
\end{cases}
\end{equation}
where $u$ is a parameter encoding the Kasner exponents. Some important key features of the classical dynamics are as follows:

${\bullet}$ {\it{Singularity}}: As $t \to 0$, the volume $V \propto e^{3\alpha} \to 0$, 
with an infinite number of Kasner transitions occurring before the singularity.

${\bullet}$ {\it{Chaos}}: The triangular billiard in the $(\beta_+,\beta_-)$ plane is hyperbolic, 
exhibiting positive Lyapunov exponents and sensitive dependence on initial conditions. Although the system exhibits deterministic chaos, in certain regimes there exist approximate constants of motion (adiabatic invariants) that remain nearly conserved over several Kasner epochs. These invariants provide a useful tool to analyze the statistical properties of the chaotic evolution.

${\bullet}$ {\it{Infinite sequence of bounces}}: The universe undergoes an endless sequence of transitions 
between different Kasner regimes, known as Mixmaster oscillations.

These remarks complete our brief review of the classical Mixmaster dynamics and will serve as a baseline for the subsequent analysis of modified scenarios. This classical picture serves as the reference against which we will compare the classical polymerized and deformed-Poisson-bracket-modified Bianchi~IX dynamics. In particular, we will examine the possible resolution or modification of the singularity, changes in the BKL transition rules and the Kasner map, suppression or alteration of chaotic behavior and modifications in the effective height and sharpness of the potential walls.

\section{Classical polymerization and deformed Poisson structures}

In this section, we first provide a brief review of the classical polymerization procedure which has been widely employed in symmetry-reduced models of quantum gravity. Then, we take a look at the idea of deformed Poisson structures, motivated from various approaches such as generalized uncertainty principles and noncommutative geometries. Finally, we highlight the motivations for combining these two ingredients in the context of the Mixmaster universe.

\subsection{Classical Polymerization}

The polymer representation arises naturally in loop quantum gravity, where the kinematical Hilbert space is constructed from holonomies of the connection rather than the connection itself. In symmetry-reduced mini-superspace models, such as the isotropic FRW universe or the Bianchi models, this leads to an effective description in which the canonical variables are modified according to the so-called ``polymerization scheme.'' 

In Schr\"{o}dinger picture of quantum mechanics, the coordinates and momentum representations are equivalent and may be easily converted to each other by a Fourier transformation.
However, in the presence of the quantum gravitational effects the space-time may take a discrete structure so that such a well-defined representations are no longer applicable. As an alternative, polymer quantization provides a suitable framework for studying these situations \cite{corichi2007polymer, ashtekar2003mathematical}. The Hilbert space of this
representation of quantum mechanics is ${\mathcal H}_{\rm poly}=L^2(R_{_d},d\mu_{_d})$, where $d\mu_{_d}$ is the Haar measure, and $R_{_d}$ denotes the real discrete line whose segments are
labeled by an extra dimension-full parameter $\mu$ such that the standard Schr\"{o}dinger picture will be recovered in the continuum limit $\mu\rightarrow\ 0$. This means that by a classical limit $\hbar\rightarrow\,0$, the polymer quantum mechanics tends to an effective $\mu$-dependent classical theory which is somehow different from the classical theory from which we have started. Such an effective theory may also be obtained directly from the standard
classical theory, without referring to the polymer quantization, by using of the Weyl operator \cite{CPR}. The process is known as {\it polymerization} with which we will deal in the rest of this paper.

According to the mentioned above form of the Hilbert space of the polymer representation of quantum mechanics, the position space (with coordinate $q$) has a discrete structure with discreteness
parameter $\mu$. Therefore, the associated momentum operator $\hat{p}$, which is the generator of the displacement, does not exist \cite{ashtekar2003mathematical, banerjee2007discreteness}. However, the Weyl exponential operator (shift operator) correspond to the discrete translation along $q$ is well defined and effectively plays the role of momentum associated to $q$
\cite{corichi2007polymer}. This allows us to utilize the Weyl operator to find an effective momentum in the semiclassical regime. So, consider a state $f(q)$, its derivative with respect to the discrete position $q$ may be approximated by means of the Weyl operator as \cite{CPR}

\begin{eqnarray}\label{FWD}
\partial_{q}f(q)\approx\frac{1}{2\mu}[f(q+\mu)-f(q-
\mu)]\hspace{2cm}\nonumber\\=\frac{1}{2\mu}\Big(
\widehat{e^{ip\mu}}-\widehat{e^{-ip\mu}}\Big)\,f(q)=
\frac{i}{\mu}\widehat{\sin(\mu p)}\,f(q),
\end{eqnarray}
and similarly the second derivative approximation will be
\begin{eqnarray}\label{SWD}
\partial_{q}^2f(q)\approx\frac{1}{\mu^2}[f(q+\mu)-2
f(q)+f(q-\mu)]\hspace{1cm}\nonumber\\=\frac{2}{\mu^2}
(\widehat{\cos(\mu p)}-1)\,f(q).\hspace{2cm}
\end{eqnarray}
Having the above approximations at hand, we define the polymerization process for the finite values of the parameter
$\mu$ as

\begin{eqnarray}\label{Polymerization}
\hat{p}\rightarrow\,\frac{1}{\mu}\widehat{\sin(\mu p)},
\hspace{1cm}\hat{p}^2\rightarrow\,\frac{2}{\mu^2}(1-
\widehat{\cos(\mu p)}).
\end{eqnarray}This replacements suggest the idea that a classical theory may be obtained via this process, but now without any attribution to the Weyl operator. This is what which is dubbed
usually as {\it classical Polymerization} in literature
\cite{corichi2007polymer, CPR}:

\begin{eqnarray}\label{PT}
q\rightarrow q,\hspace{1.5cm}p\rightarrow\frac{ \sin(\mu
p)}{\mu},\hspace{5mm}p^2\rightarrow
\frac{2}{\mu^2}\left[1-\cos(\mu p)\right],
\end{eqnarray}where now $(q,p)$ are a pair of classical phase space variables. Equivalently, one may view this replacement as originating from the underlying compactification of the momentum space in the polymer Hilbert space. For small values of $\mu p$, the standard momentum is recovered, while at higher energies the sinusoidal modification leads to bounded functions which mimic certain quantum geometric effects. Hence, by applying the transformation (\ref{PT}) to the Hamiltonian of a classical system we get its classical polymerized counterpart. A glance at (\ref{PT}) shows that the momentum is periodic and varies in a bounded interval as $p\in[-\frac{\pi}{\mu},+\frac{\pi}{ \mu})$. In the limit
$\mu\rightarrow\,0$, one recovers the usual range for the canonical momentum $p\in(-\infty,+\infty)$. Therefore, the polymerized momentum is compactified and topology of the momentum
sector of the phase space is $S^1$ rather than the usual $\mathbb{R}$ \cite{NaturalCutoff}. Our set-up to explain the classical polymerization of a dynamical system is now complete. This procedure has been applied to various cosmological models as an effective way to incorporate some aspects of quantum gravitational effects without relying on the full quantum formalism. In particular, when applied to homogeneous cosmologies, the resulting effective dynamics often exhibit a bounce replacing the initial big-bang singularity \cite{bojowald2001absence}.

At the level of the Hamiltonian dynamics, polymerization introduces non-linear corrections that can significantly alter the behavior of the system. For example, in cosmological models and black hole physics, such modifications often result in the resolution of the initial singularity by replacing it with a non-singular bounce \cite{qc pol} and \cite{pol bl}. In anisotropic models, such as Bianchi I, the polymerized dynamics lead to bounded expansion rates and modified shear terms, thereby controlling the approach to singularities. 

This effective scheme has been applied to a variety of models including FRW, Bianchi I, and even black hole interiors, offering a tractable framework to explore semiclassical effects of LQG in a classical Hamiltonian setting.

\subsection{Deformed Poisson brackets}

Another line of investigation in quantum gravity and related areas considers deformations of the fundamental Poisson algebra. Instead of assuming the canonical structure
\begin{equation}
\{q_i, p_j\} = \delta_{ij}, \qquad \{q_i, q_j\} = \{p_i, p_j\} = 0,
\end{equation}
one postulates modified brackets of the form
\begin{equation}
\{q_i, p_j\} = \delta_{ij} \, f(p,q), 
\end{equation}
where $f(p,q)$ encodes the deformation and typically reduces to unity in the classical limit. Such deformations arise in several contexts: (i) The \textit{generalized uncertainty principle} (GUP), motivated from string theory and black hole physics, suggests a modification of the canonical algebra that incorporates a minimal length scale \cite{kempf1995, GUP Noz}, (ii) \textit{Noncommutative geometries}, such as Snyder space or $\kappa$-Minkowski spacetime, lead to non-trivial commutators among spatial coordinates and their conjugate momenta, which at the classical level correspond to deformed Poisson brackets \cite{snyder}, (iii) {\it Deformed symplectic structures} also appear in certain effective models of loop quantum cosmology, where holonomy and inverse triad corrections can be encoded in modified algebraic relations \cite{kon}. 

The general consequence of such deformations is that the phase space acquires a non-trivial structure, which in turn affects the Hamiltonian evolution. For instance, in anisotropic cosmological models, a deformed algebra can lead to modifications in the Kasner exponents, alter the chaotic dynamics of the Mixmaster universe, or generate bounds on curvature invariants \cite{vakilibianki}. 

The motivation for considering polymerization together with deformed Poisson structures is twofold. On the one hand, polymerization captures essential aspects of loop quantization by introducing non-linear trigonometric modifications to the canonical variables, effectively regularizing certain divergences and leading to bounded dynamics. On the other hand, deformed Poisson brackets provide a more general framework in which non-classical effects—arising from minimal length scales, noncommutativity, or modified symplectic geometry—are consistently encoded at the level of the algebra.  

In the context of the Mixmaster (Bianchi IX) universe, the interplay between these two ingredients becomes particularly relevant. The classical Mixmaster dynamics is known to exhibit chaotic oscillations in the approach to the singularity. Understanding how quantum gravitational effects might tame or modify this chaotic behavior has been a long-standing problem. By implementing polymerization together with a specific choice of deformed Poisson brackets, we aim to analyze whether the resulting effective dynamics can suppress or alter the chaotic regime, and if so, under what conditions. This combined approach not only provides new insights into the resolution of cosmological singularities but also sheds light on the broader question of how semiclassical modifications of the symplectic structure affect the dynamics of anisotropic universes. 

In the present work, we shall keep the above discussion at a general level and only assume that both types of modifications ---namely, classical polymerization and deformed Poisson brackets--- may play a role in the effective dynamics of the Bianchi IX universe. The precise choice of deformation function $f(q,p)$ will be specified in the following sections, once the Hamiltonian framework of the model is established. This allows us to set up the formalism in full generality before focusing on a particular class of deformations.

\section{Polymerized Hamiltonian and equations of motion with deformed Poisson brackets}

We adopt the Misner variables $(\alpha,\beta_+,\beta_-)$ with their canonical momentum conjugates $(p_\alpha,p_+,p_-)$ and keep the classical Bianchi~IX potential $V(\beta_+,\beta_-)$. In units and lapse matching our classical section, the classical constraint is given by equation (\ref{Hamil}). Following the standard polymer prescription, we replace each momentum by a bounded trigonometric function
\begin{equation}
p_i \;\longmapsto\; \Pi_i(p_i) \;\equiv\; \frac{1}{\mu_i}\,\sin(\mu_i p_i), 
\qquad i\in\{\alpha,+,-\},
\label{eq:poly-map}
\end{equation}
with polymer scales $\mu_i$ (not necessarily equal) that reduce to the classical theory as $\mu_i\!\to\!0$. The polymerized Hamiltonian constraint is then
\begin{equation}
\mathcal{H}_{\rm poly} \;=\; -\,\Pi_\alpha^2 \;+\; \Pi_+^2 \;+\; \Pi_-^2 \;+\; e^{4\alpha}\,V(\beta_+,\beta_-)\;\approx 0.
\label{eq:H_poly}
\end{equation}
By construction, $\Pi_i$ are bounded, which effectively regularizes the kinetic sector and can prevent unbounded growth of curvature invariants in the effective dynamics. The classical limit is recovered via $\sin(\mu_i p_i)\!\sim\!\mu_i p_i$. It should be emphasized that one may allow anisotropic polymer scales $\mu_\alpha$, $\mu_+$ and $\mu_-$ to track distinct UV sensitivities of volume and shear degrees of freedom. Note also that nothing is assumed about the potential term $V$, beyond its standard Bianchi~IX form. All quantum--gravity–inspired corrections are encoded in the kinetic sector (polymerization) and/or in the symplectic structure (deformation) below.

To keep the discussion general, we allow the fundamental brackets to be deformed in a diagonal minisuperspace form

\begin{equation}
\{\alpha,p_\alpha\}=F_\alpha(\alpha,\beta_\pm,p_\alpha,p_\pm),\quad
\{\beta_+,p_+\}=F_+(\alpha,\beta_\pm,p_\alpha,p_\pm),\quad
\{\beta_-,p_-\}=F_-(\alpha,\beta_\pm,p_\alpha,p_\pm),
\label{eq:PB-deformed}
\end{equation}
with all other brackets vanishing.\footnote{This choice preserves the minisuperspace block structure and is sufficiently general for many deformations (GUP-like, Snyder-like, effective LQC-inspired). The specific functional forms $F_i$ will be fixed later. In the classical limit $F_i\to 1$.}
Given that $\Pi_i=\Pi_i(p_i)$, the chain rule implies

\begin{equation}
\{q_i,\Pi_i\} \;=\; F_i\,\frac{\partial \Pi_i}{\partial p_i} 
\;=\; F_i \,\cos(\mu_i p_i), 
\qquad q_i\in\{\alpha,\beta_+,\beta_-\}.
\label{eq:PB-q-Pi}
\end{equation}
Now, the Hamilton's equations with the deformed brackets read

\begin{align}
\dot{\alpha} \;&=\; \{\alpha,\mathcal{H}_{\rm poly}\}
\;=\; \{\alpha,\Pi_\alpha\}\,\frac{\partial \mathcal{H}_{\rm poly}}{\partial \Pi_\alpha}
\;=\; F_\alpha\,\cos(\mu_\alpha p_\alpha)\,(-2\Pi_\alpha),
\label{eq:alpha-dot}\\[3pt]
\dot{\beta}_\pm \;&=\; \{\beta_\pm,\mathcal{H}_{\rm poly}\}
\;=\; \{\beta_\pm,\Pi_\pm\}\,\frac{\partial \mathcal{H}_{\rm poly}}{\partial \Pi_\pm}
\;=\; F_\pm\,\cos(\mu_\pm p_\pm)\,(2\Pi_\pm),
\label{eq:beta-dot}\\[3pt]
\dot{p}_\alpha \;&=\; \{p_\alpha,\mathcal{H}_{\rm poly}\}
\;=\; -\,F_\alpha\,\frac{\partial \mathcal{H}_{\rm poly}}{\partial \alpha}
\;=\; -\,4\,F_\alpha\,e^{4\alpha}\,V(\beta_+,\beta_-),
\label{eq:pa-dot}\\[3pt]
\dot{p}_\pm \;&=\; \{p_\pm,\mathcal{H}_{\rm poly}\}
\;=\; -\,F_\pm\,\frac{\partial \mathcal{H}_{\rm poly}}{\partial \beta_\pm}
\;=\; -\,F_\pm\,e^{4\alpha}\,\frac{\partial V}{\partial \beta_\pm}.
\label{eq:pp-dot}
\end{align}
It is worth noting that in deriving the equations of motion we have 
set the lapse function to unity, $N=1$. This choice corresponds to 
using the so--called ``Misner time'' parametrization, or equivalently 
introducing a new time variable $\tau$ defined by $d\tau = N\,dt$. 
In this parametrization the Hamiltonian equations read $\dot{q}=\{q,\mathcal{H}\}$ and $\dot{p}=\{p,\mathcal{H}\},$ which is the convention we adopt throughout this work. 
Restoring a general lapse $N$ is straightforward and amounts to multiplying the right--hand side of all dynamical equations by $N$. 
This does not alter the constraint surface $\mathcal{H}\approx 0$, but simply corresponds to a reparametrization of the time variable. Together with the constraint \eqref{eq:H_poly}, these equations define the effective dynamics. In the classical limit $\mu_i\!\to\!0$ and $F_i\!\to\!1$, one recovers $\dot{\alpha}=-2p_\alpha$, $\dot{\beta}_\pm=2p_\pm$, and the classical force terms.

Equations \eqref{eq:alpha-dot}–\eqref{eq:pp-dot} show that polymerization suppresses large shear momenta via the bounded $\Pi_\pm$, while deformation factors $F_\pm$ rescale the effective ``velocities'' in anisotropy space. Near potential walls, the force terms $\propto e^{4\alpha}\partial_{\beta_\pm}V$ dominate; polymer/deformation may modify reflection laws (BKL map) through the prefactors $F_\pm \cos(\mu_\pm p_\pm)$.

From \eqref{eq:alpha-dot}, turning points in the volume variable occur when either $\Pi_\alpha=0$ (i.e.\ $\sin(\mu_\alpha p_\alpha)=0$) or $\cos(\mu_\alpha p_\alpha)=0$. The latter corresponds to extrema of $\Pi_\alpha(p_\alpha)$ and is a purely polymer effect. Whether these turning points realize a true bounce depends on the constraint \eqref{eq:H_poly} and the sign of $\dot{p}_\alpha$ in \eqref{eq:pa-dot}; both are sensitive to $F_\alpha$ and to the potential $V$ (wall encounters).

In the classical limit for which $\mu_i\!\to\!0$ and $F_i\!\to\!1$, $\Pi_i\!\to\!p_i$ and \eqref{eq:H_poly}–\eqref{eq:pp-dot} reduce to the standard equations of motion.
Also, for the small-deformation expansion $F_i=1+\delta F_i$, with $|\delta F_i|\ll 1$ and $\mu_i p_i\ll 1$, we have

\begin{equation}
\Pi_i = p_i - \tfrac{1}{6}\mu_i^2 p_i^3 + \mathcal{O}(\mu_i^4), 
\qquad 
\cos(\mu_i p_i)=1 - \tfrac{1}{2}\mu_i^2 p_i^2 + \mathcal{O}(\mu_i^4),
\end{equation}
which yields a controlled perturbative scheme for tracing polymer/deformation impacts on Kasner transitions and Lyapunov indicators.

We now specify the deformation functions entering the Poisson brackets
\(
\{\alpha,p_\alpha\}=F_\alpha,\ 
\{\beta_+,p_+\}=F_+,\ 
\{\beta_-,p_-\}=F_-,
\)
with all other brackets vanishing. To ensure the Jacobi identities in minisuperspace, we adopt diagonal deformations where each $F_i$ depends only on its own canonical pair or momenta.\footnote{With the bracket structure chosen in \eqref{eq:PB-deformed}, the Jacobi identities are automatically satisfied if $F_i$ are functions of $(q_i,p_i)$ alone (no cross dependence). Cross-coupled choices can also be considered but require additional consistency conditions.} 
We introduce four representative families:

${\bullet}$ {\it{GUP-like, momentum-diagonal (baseline)}}: 

\begin{equation}
F_\alpha \;=\; 1+\gamma_\alpha\,p_\alpha^2,\qquad
F_\pm \;=\; 1+\gamma_\pm\,p_\pm^2,
\label{eq:F-GUP-diag}
\end{equation}
with small, positive deformation parameters $\gamma_i\ge 0$. 
This class preserves analyticity, yields a controlled small-deformation expansion, and keeps the equations of motion algebraically simple

\begin{equation}
\dot{\alpha} = -2\,\bigl(1+\gamma_\alpha p_\alpha^2\bigr)\,\cos(\mu_\alpha p_\alpha)\,\Pi_\alpha,
\quad
\dot{\beta}_\pm = 2\,\bigl(1+\gamma_\pm p_\pm^2\bigr)\,\cos(\mu_\pm p_\pm)\,\Pi_\pm,
\end{equation}
with force terms scaled by the same $F_i$ as in Eqs.~\eqref{eq:pa-dot}–\eqref{eq:pp-dot}.
At leading order ($\mu_i p_i\ll 1$, $\gamma_i p_i^2\ll 1$) one has

\begin{equation}
\dot{\alpha} = -2 p_\alpha\left[1+\gamma_\alpha p_\alpha^2-\tfrac12\mu_\alpha^2 p_\alpha^2+\mathcal{O}(\mu^4,\gamma\mu^2,\gamma^2)\right],
\end{equation}
and similarly for $\dot{\beta}_\pm$. This facilitates perturbative control over modifications of Kasner epochs and reflection laws.

${\bullet}$ {\it{Isotropic-in-momentum (total-norm) deformation}}: 

\begin{equation}
F_\alpha = F_+ = F_- \equiv 1+\Gamma\,\bigl(p_\alpha^2+p_+^2+p_-^2\bigr).
\label{eq:F-isotropic}
\end{equation}
This choice treats the three momenta on equal footing and may be motivated by symmetry considerations. It couples the three directions through a single prefactor, modifying speeds in minisuperspace uniformly. Analytically it is still tractable, though reflection rules acquire a mild dependence on the total momentum norm.

${\bullet}$ {\it{Polymer-compatible algebraic dressing}}: 

\begin{equation}
F_i \;=\; \cos(\mu_i p_i),
\qquad i\in\{\alpha,+,-\}.
\label{eq:F-cos}
\end{equation}
This LQC-inspired ansatz aligns the symplectic deformation with the polymer map $\Pi_i'=\cos(\mu_i p_i)$, yielding

\begin{equation}
\dot{q}_i \;=\; 2\,\cos^2(\mu_i p_i)\,(\pm \Pi_i), 
\end{equation}
i.e.\ a \emph{double} cosine suppression of velocities. This prescription modifies the effective Hamiltonian in such a way that the bounce
mechanism and the rules for BKL reflections are altered. It often results in compactification of the momentum space and emphasizes periodic dynamics in $p_i$. It strongly tames dynamics near $\mu_i p_i\simeq \frac{\pi}{2}$ and can significantly modify billiard reflections. Care is required to avoid zero crossings of $F_i$ (signature-like effects).

${\bullet}$ {\it{Standard LQC-inspired polymerization}}: The most common choice, directly
motivated by loop quantum cosmology, is

\begin{equation}
F_i(p_i)=\frac{\sin(\mu_i p_i)}{\mu_i},
\end{equation}
where $\mu_i$ are polymer scales associated with each degree of freedom. This modification reduces to the classical variable $p_i$ in the limit $\mu_i\rightarrow 0$, and leads to a natural
boundedness of the effective curvature terms, preventing singularities and ensuring the possibility of a bounce.

For analytic transparency, we adopt the diagonal GUP-like deformation \eqref{eq:F-GUP-diag} as the \emph{baseline} in the main text, with $\gamma_i\ge 0$ small. With \eqref{eq:F-GUP-diag}, the effective ``velocities'' in minisuperspace are rescaled by $(1+\gamma_i p_i^2)\cos(\mu_i p_i)$, while forces are scaled by $(1+\gamma_i p_i^2)$. 
Hence, polymerization bounds the momenta via $\Pi_i$, and the GUP-like factors further suppress large-$|p_i|$ flows without introducing zeros in $F_i$. 
Turning points in $\alpha$ arise when $\Pi_\alpha=0$ or $\cos(\mu_\alpha p_\alpha)=0$ (polymer-induced), while the GUP factor adjusts the bounce condition quantitatively but not qualitatively.
Classical dynamics is recovered for $\mu_i\to 0$ and $\gamma_i\to 0$. To avoid superluminal parametrization effects and preserve monotonic time reparametrization, we assume $F_i>0$ along physical trajectories, guaranteed for \eqref{eq:F-GUP-diag} with $\gamma_i\ge 0$. Small-deformation expansions in $(\mu_i,\gamma_i)$ will be used when assessing Kasner transitions and Lyapunov indicators.

As mentioned before, in the classical Mixmaster framework, the dynamics near the singularity are governed by a sequence of successive reflections (the BKL map) on the \((\beta_{+},\beta_{-})\) plane. These reflections generate chaotic and unpredictable behavior, which is the hallmark of the classical Mixmaster model. With polymerization, two major qualitative modifications are expected:  
First: {\it Suppression of the effective energy close to the singularity,} because of the built--in cut--off scale in the sinusoidal/cosine substitution functions, divergent energy values in the canonical variables no longer occur, and an upper bound for the Hamiltonian energy emerges. Second: {\it Modification of the BKL reflection law:} in the sence that the intensity and frequency of the reflections are reduced. In other words, the system still experiences BKL--type oscillations, but in a limited range and with a ``softened reflection.''  

Our aim is to show that the chosen deformation functions \(F_i(p_i)\) together with the classical polymerization of the model can lead to distinct corrections in the reflection pattern. 
Comparing these choices allows us to address the key question: \emph{Does combined GUP+polymer completely remove the chaotic behavior, or does it merely transform it into a ``controlled chaos''?}

\section{GUP-deformed perturbative effective dynamics}

We now consider the diagonal GUP--type deformation of the Poisson structure
\begin{equation}
\{\alpha,p_\alpha\}=1+\gamma_\alpha p_\alpha^2,\qquad
\{\beta_\pm,p_\pm\}=1+\gamma_\pm p_\pm^2,
\label{eq:GUP-PB}
\end{equation}
with the momenta remaining linear, \(\Pi_i=p_i\). Here \(\gamma_i\) are small deformation parameters (with dimensions inverse to \(p_i^2\) in our units) and in the classical limit \(\gamma_i\to 0\), the canonical algebra is recovered. To avoid sign changes in the effective time flow one typically assumes \(1+\gamma_i p_i^2>0\), along physical trajectories; for \(\gamma_i\ge 0\) this condition is automatically satisfied. Also, for practical computations we will use small values of \(\gamma_i\) (e.g.\ \(\gamma\lesssim 10^{-2}\) in appropriate units) to remain within the perturbative regime. The exact admissible range must be tested numerically depending on initial conditions  and the chosen lapse (gauge) function. With the undeformed Hamiltonian (\ref{Hamil}) the deformed Poisson brackets \eqref{eq:GUP-PB} yield the following equations of motion

\begin{align}
\dot\alpha &= \{\alpha,\mathcal{H}_{\rm cl}\} = -2\,(1+\gamma_\alpha p_\alpha^2)\,p_\alpha,
\label{eq:alpha-dot-gup}\\[4pt]
\dot\beta_\pm &= \{\beta_\pm,\mathcal{H}_{\rm cl}\} = 2\,(1+\gamma_\pm p_\pm^2)\,p_\pm,
\label{eq:beta-dot-gup}\\[4pt]
\dot p_\alpha &= \{p_\alpha,\mathcal{H}_{\rm cl}\} = -4\,(1+\gamma_\alpha p_\alpha^2)\,e^{4\alpha}V(\beta_+,\beta_-),
\label{eq:pa-dot-gup}\\[4pt]
\dot p_\pm &= \{p_\pm,\mathcal{H}_{\rm cl}\} = -(1+\gamma_\pm p_\pm^2)\,e^{4\alpha}\frac{\partial V}{\partial\beta_\pm}.
\label{eq:pp-dot-gup}
\end{align}
For perturbative analysis we assume \(|\gamma_i p_i^2|\ll 1\). Expanding to first order in \(\gamma\) gives
\begin{align}
\dot\alpha &= -2p_\alpha - 2\gamma_\alpha p_\alpha^3 + \mathcal{O}(\gamma^2), \\
\dot\beta_\pm &= 2p_\pm + 2\gamma_\pm p_\pm^3 + \mathcal{O}(\gamma^2), \\
\dot p_\alpha &= -4e^{4\alpha}V(\beta_+,\beta_-) -4\gamma_\alpha p_\alpha^2 e^{4\alpha}V(\beta_+,\beta_-) + \mathcal{O}(\gamma^2), \\
\dot p_\pm &= -e^{4\alpha}\partial_{\beta_\pm}V - \gamma_\pm p_\pm^2 e^{4\alpha}\partial_{\beta_\pm}V + \mathcal{O}(\gamma^2).
\end{align}
Unlike the polymer case, here the momenta \(p_i\) are not intrinsically bounded by the kinematic map. The necessary condition for a turning point of the volume variable is \(\dot\alpha=0\), which (to first order in \(\gamma\)) implies
\begin{equation}
-2p_\alpha - 2\gamma_\alpha p_\alpha^3 \approx 0
\quad\Longrightarrow\quad
p_\alpha\bigl(1+\gamma_\alpha p_\alpha^2\bigr)\approx 0.
\end{equation}
Hence the trivial root \(p_\alpha=0\), is the only perturbative solution; a nontrivial turning point requires dynamics that drive \(p_\alpha\) through zero (i.e.\ a sign change), which depends on the force equation \(\dot p_\alpha\). Because \(\dot p_\alpha\) carries the factor \((1+\gamma_\alpha p_\alpha^2)\,e^{4\alpha}V\), GUP alone does not generically guarantee a bounce: momentum can continue to grow large in magnitude unless other effects (e.g.\ strong potential terms or polymer--type bounds) intervene. In short, \emph{GUP-type deformation typically weakens the approach to the singularity (by suppressing effective velocities) but does not by itself ensure a non-singular bounce}.

In the BKL picture, reflections occur when the potential wall exerts a dominant impulse on the anisotropy momenta. To leading perturbative order in \(\gamma\), the incoming/outgoing anisotropy velocities are rescaled by the factor \((1+\gamma_\pm p_\pm^2)\). Consequently the Kasner parameter map receives small \(\gamma\)-dependent corrections of the schematic form
\begin{equation}
u_{n+1} \;\approx\; \mathcal{G}_{\rm cl}(u_n) \;+\; \Delta u(\gamma),
\qquad
\Delta u(\gamma) \propto \gamma\,u_n^3 + \mathcal{O}(\gamma^2),
\end{equation}
where \(\mathcal{G}_{\rm cl}\) denotes the classical BKL map. These corrections are perturbative and typically softer than the nonperturbative modifications induced by polymerization. Therefore, for small \(\gamma\) one expects a gradual suppression of chaotic indicators (decrease of effective Lyapunov exponent) rather than an abrupt structural change of the billiard dynamics.

To evaluate the perturbative correction to the BKL map for the GUP--type deformation, we derive a first--order correction in the diagonal GUP parameter $\gamma$ to the 
classical BKL reflection law. Our derivation follows the standard ``impulse'' (thin-wall) approximation: the interaction with one potential wall is assumed to 
be short in the chosen time parameter and dominated by the wall normal component of the force. We denote by $n$ the normal direction to the considered wall in the 
$(\beta_+,\beta_-)$ plane and by $t$ the tangential direction. To do this, let the anisotropy canonical pair in the wall-adapted basis be $(\beta_n,p_n)$ and 
$(\beta_t,p_t)$ (with $p_n$ the normal momentum). The classical elastic reflection 
off a steep exponential wall reverses the normal momentum

\begin{equation}
p_n^{\rm (out)} = -\,p_n^{\rm (in)}, \qquad p_t^{\rm (out)} = p_t^{\rm (in)}.
\end{equation}
The Kasner parameter \(u\) is a function of the anisotropy momenta (or equivalently 
of the Kasner exponents); for the present perturbative purpose we treat \(u\) as a 
smooth function of the incoming momenta, \(u = u(p_n,p_t,\dots)\), and linearize 
its change under the small correction to the momentum impulse.

With the diagonal GUP deformation

\begin{equation}
\{\beta_n,p_n\}=1+\gamma p_n^2,
\end{equation}
the equation of motion for $p_n$ reads 

\begin{equation}
\dot p_n \;=\; -\,(1+\gamma p_n^2)\, e^{4\alpha}\,\frac{\partial V}{\partial \beta_n}.
\end{equation}
Integrating across the short interaction time for the $n$--th wall (from $t_{\rm in}$ to $t_{\rm out}$),

\begin{equation}
\Delta p_n \equiv p_n^{\rm(out)}-p_n^{\rm(in)}
= -\int_{t_{\rm in}}^{t_{\rm out}} (1+\gamma p_n^2)\, e^{4\alpha}\, \partial_{\beta_n}V \,dt.
\end{equation}
Split the integral into the classical part plus the $\gamma$--correction

\begin{equation}
\Delta p_n = \Delta p_n^{(0)} + \gamma\,\Delta p_n^{(1)} + \mathcal{O}(\gamma^2),
\end{equation}
where

\begin{equation}
\Delta p_n^{(0)} = -\int_{t_{\rm in}}^{t_{\rm out}} e^{4\alpha}\, \partial_{\beta_n}V \,dt,
\qquad
\Delta p_n^{(1)} = -\int_{t_{\rm in}}^{t_{\rm out}} p_n^2\, e^{4\alpha}\, \partial_{\beta_n}V \,dt.
\end{equation}
In the thin-wall (impulse) limit the classical integral yields the elastic flip

\begin{equation}
\Delta p_n^{(0)} = -2 p_n^{\rm(in)}.
\end{equation}
(The classical result follows from energy--momentum matching across the brief impulse; exact derivations in the Mixmaster literature show the normal component reverses sign and doubles in magnitude). The first--order correction is

\begin{equation}
\Delta p_n = -2 p_n^{\rm(in)} + \gamma\,\Delta p_n^{(1)} + \mathcal{O}(\gamma^2).
\end{equation}
The correction integral \(\Delta p_n^{(1)}\) depends on the interaction profile and on the 
time--dependence of \(p_n\) during the collision. Under the usual thin--wall approximation 
one may replace $p_n(t)$ inside the integral by its incoming value \(p_n^{\rm(in)}\) to 
leading order, hence

\begin{equation}
\Delta p_n^{(1)} \approx -\,\bigl(p_n^{\rm(in)}\bigr)^2
\int_{t_{\rm in}}^{t_{\rm out}} e^{4\alpha}\,\partial_{\beta_n}V \,dt
= \bigl(p_n^{\rm(in)}\bigr)^2 \,\bigl(-\Delta p_n^{(0)}\bigr)
= 2 \,\bigl(p_n^{\rm(in)}\bigr)^3,
\end{equation}
where we used \(-\int e^{4\alpha}\partial_{\beta_n}V dt = 2 p_n^{\rm(in)}\), from the classical impulse.
Therefore, to first order in \(\gamma\),

\begin{equation}
\Delta p_n \;\approx\; -2 p_n^{\rm(in)} \;+\; 2\gamma \,\bigl(p_n^{\rm(in)}\bigr)^3.
\end{equation}
Now, we may compute the induced change in the Kasner parameter \(u\). Let \(u\) be expressed (locally) as a smooth function of the anisotropy momenta; 
for small changes we can linearize

\begin{equation}
\Delta u \;=\; u\bigl(p_n^{\rm(out)},p_t^{\rm(out)}\bigr) - u\bigl(p_n^{\rm(in)},p_t^{\rm(in)}\bigr)
\approx \frac{\partial u}{\partial p_n}\Big|_{\rm in}\, \Delta p_n
+ \frac{\partial u}{\partial p_t}\Big|_{\rm in}\, \Delta p_t.
\end{equation}
Classically \(\Delta p_t=0\) and \(\Delta p_n^{(0)}=-2p_n^{\rm(in)}\), hence the classical change is

\begin{equation}
\Delta u^{(0)} = \frac{\partial u}{\partial p_n}\Big|_{\rm in}\,(-2p_n^{\rm(in)}).
\end{equation}
Including the GUP correction

\begin{equation}
\Delta u \approx \Delta u^{(0)} \;+\; \gamma \,\frac{\partial u}{\partial p_n}\Big|_{\rm in}
\left(2 (p_n^{\rm(in)})^3\right) + \mathcal{O}(\gamma^2).
\end{equation}
Collecting the above expressions we obtain the schematic linear correction formula

\begin{equation}
u_{n+1} \;=\; \mathcal{G}_{\rm cl}(u_n) \;+\; \gamma \; \mathcal{S}(u_n) \;+\; \mathcal{O}(\gamma^2),
\label{eq:Delta-u-gamma}
\end{equation}
where the first--order correction
\(\mathcal{S}(u_n)\) can be written as

\begin{equation}
\mathcal{S}(u_n) \;=\; 2\,\left[p_n^{\rm(in)}(u_n)\right]^3 \;\frac{\partial u}{\partial p_n}\Big|_{\rm in}.
\label{eq:S-of-u}
\end{equation}
Equations \eqref{eq:Delta-u-gamma} and \eqref{eq:S-of-u} are exact within the thin--wall 
and ``frozen-momentum'' approximations used above; the entire dependence on the wall 
profile and on the chosen parametrization is encoded in the functions \(p_n^{\rm(in)}(u)\) 
and \(\partial u/\partial p_n\), both of which can be expressed in terms of Kasner 
parameters or computed numerically for a given initial condition.

The result shows that the GUP correction is cubic in the incoming normal momentum, hence it is more important for collisions with large incoming \(p_n\). This is consistent with 
the physical intuition that GUP corrections are relevant at high momenta. To obtain an explicit function \(\mathcal{S}(u)\) one may (i) use the standard Kasner parametrization to express \(p_n^{\rm(in)}\) and \(\partial u/\partial p_n\) analytically in terms of \(u\), or (ii) evaluate these quantities numerically from the classical (unperturbed) approach-to-wall trajectory and then compute \(\mathcal{S}(u)\). The thin--wall approximation and the replacement \(p_n(t)\!\mapsto\!p_n^{\rm(in)}\), inside the integral are controlled approximations when the collision time is short compared with the typical time scale of variation of $p_n$. If a more accurate coefficient is desired, one must compute the integral \(\Delta p_n^{(1)} = -\int p_n^2(t) e^{4\alpha(t)}\partial_{\beta_n}V\,dt\), using the actual time profile \(p_n(t)\) during the interaction. This is straightforward numerically.

We now obtain an explicit analytic expression for the first--order correction 
function $\mathcal S(u)$ appearing in the GUP correction to the BKL map,
cf.\ Eq.~\eqref{eq:Delta-u-gamma}. We specialize to the reflection against the 
wall primarily associated with the $\beta_+$ direction (the calculation for 
the other walls is analogous up to cyclic permutations). The Kasner exponents may be parametrized by the classical BKL parameter $u\ge 1$ as

\begin{equation}
p_1(u) = -\frac{u}{1+u+u^2},\qquad
p_2(u) = \frac{1+u}{1+u+u^2},\qquad
p_3(u) = \frac{u(1+u)}{1+u+u^2}.
\label{eq:kasner-u}
\end{equation}
These satisfy $p_1+p_2+p_3=1$ and $p_1^2+p_2^2+p_3^2=1$. In the Misner variables the anisotropy canonical momenta are linear combinations 
of the Kasner exponents. A convenient (and standard) choice for the momentum associated with the $\beta_+$ direction is

\begin{equation}
p_+ \;=\; \frac{1}{2}\bigl(p_1 + p_2 - 2 p_3\bigr).
\label{eq:pplus-def}
\end{equation}
Inserting \eqref{eq:kasner-u} into \eqref{eq:pplus-def} yields a closed form

\begin{equation}
p_+ (u) \;=\; \frac{1 - 2u - 2u^2}{2(1+u+u^2)}.
\label{eq:pplus-u}
\end{equation}
For the other walls one proceeds by cyclic permutation of the indices. We may now compute the derivative of \eqref{eq:pplus-u} with respect to $u$.
Let $D(u)=1+u+u^2$ and $N(u)=1-2u-2u^2$, so $p_+(u)=N/(2D)$. Then

\begin{equation}
\frac{dp_+}{du} \;=\; \frac{1}{2}\,\frac{N'(u)D(u)-N(u)D'(u)}{D(u)^2}.
\end{equation}
Explicitly

\begin{equation}
N'(u) = -2-4u,\qquad D'(u) = 1+2u,
\end{equation}
hence after simplifying

\begin{equation}
\frac{dp_+}{du}
\;=\; \frac{ -2-4u }{2D} \;-\; \frac{N(u)(1+2u)}{2D^2}
\;=\; \frac{-2-4u}{2(1+u+u^2)} \;-\; \frac{(1-2u-2u^2)(1+2u)}{2(1+u+u^2)^2}.
\label{eq:dpplus-du}
\end{equation}
One may combine the terms over the common denominator $2(1+u+u^2)^2$ to get
an algebraic rational expression; for practical purposes either form is acceptable. To evaluate an expression for $\mathcal S(u)$, recalling Eq.~\eqref{eq:S-of-u}

\begin{equation}
\mathcal S(u) \;=\; 2 \,\left[p_n^{\rm(in)}(u)\right]^3 \;\frac{\partial u}{\partial p_n}\Big|_{\rm in}
\;=\; 2 \,\frac{\left[p_+(u)\right]^3}{\dfrac{dp_+}{du}} ,
\end{equation}
where we used $\partial u/\partial p_+ = 1/(dp_+/du)$. Substituting 
\eqref{eq:pplus-u} and \eqref{eq:dpplus-du} we obtain an explicit rational function

\begin{equation}\label{eq:Su-final}
\mathcal S(u)=\frac{(2u^{2}+2u-1)^{3}}{6(2u+1)(u^{2}+u+1)}.
\end{equation}
Equation \eqref{eq:Su-final} provides an explicit analytic form for the first--order GUP correction to the BKL map for collisions with the $\beta_+$ wall. The corrections for the other walls follow by cyclic permutation of the Kasner exponents (i.e.\ by the substitutions $u\mapsto$ the corresponding transformed parameter).
The overall sign and numerical prefactor are consistent with the thin--wall, frozen-momentum approximation used in the derivation; a more accurate coefficient can be obtained by evaluating the integral \(\Delta p_n^{(1)}\) with the actual time--profile \(p_n(t)\).
 
\section{Polymerized (and combined GUP+polymer) perturbative effective dynamics}
We now specialize to the case of the LQC--inspired polymerization scheme, where 

\begin{equation}\label{Polymer}
\Pi_i(p_i) \;=\; \frac{\sin(\mu_i p_i)}{\mu_i}, 
\qquad F_i(p_i)=1,
\end{equation}
with $\mu_i$, as before, denoting the polymer scale associated with each canonical pair. 
In order to make analytic progress, it is convenient to expand $\Pi_i$ in a power 
series around the classical regime $\mu_i\to 0$:
\begin{equation}
\Pi_i(p_i) = p_i - \frac{\mu_i^2}{6}\,p_i^3 + \mathcal{O}(\mu_i^4).
\label{eq:poly_expansion}
\end{equation}
Inserting \eqref{eq:poly_expansion} into the polymer Hamiltonian leads to the effective Hamiltonian

\begin{equation}
\mathcal{H}_{\rm eff} \;=\; 
-\left(p_\alpha^2 - p_+^2 - p_-^2 \right) 
+ \frac{1}{6}\,\mu_\alpha^2\,p_\alpha^4 
- \frac{1}{6}\,\mu_+^2\,p_+^4 
- \frac{1}{6}\,\mu_-^2\,p_-^4 
+ V(\alpha,\beta_+,\beta_-)
+ \mathcal{O}(\mu^4).
\end{equation}
The corresponding equations of motion take the schematic form
\begin{align}
\dot\alpha &= -\frac{\partial \mathcal{H}_{\rm eff}}{\partial p_\alpha} 
= 2\,p_\alpha - \frac{2}{3}\,\mu_\alpha^2 p_\alpha^3 
+ \mathcal{O}(\mu^4), \\[1ex]
\dot p_\alpha &= -\frac{\partial \mathcal{H}_{\rm eff}}{\partial \alpha}
= -\,\frac{\partial V}{\partial \alpha}, \\[1ex]
\dot\beta_\pm &= \frac{\partial \mathcal{H}_{\rm eff}}{\partial p_\pm} 
= -2\,p_\pm + \frac{2}{3}\,\mu_\pm^2 p_\pm^3 
+ \mathcal{O}(\mu^4), \\[1ex]
\dot p_\pm &= -\frac{\partial \mathcal{H}_{\rm eff}}{\partial \beta_\pm}
= -\,\frac{\partial V}{\partial \beta_\pm}.
\end{align}
In the classical model, the variable $\alpha$ monotonically decreases toward 
$-\infty$ as the singularity is approached. With polymer corrections, however, 
$\dot\alpha$ receives a negative cubic term in $p_\alpha$, which allows for the 
possibility of a turning point where $\dot\alpha=0$. At leading order, the 
bounce occurs when
\begin{equation}
p_\alpha^2 \;\approx\; \frac{3}{\mu_\alpha^2}.
\label{eq:bounce_condition}
\end{equation}
At this point the contraction halts and $\dot\alpha$ changes sign, signaling a 
non--singular bounce in the volume variable.

The anisotropy variables $(\beta_+,\beta_-)$ undergo BKL reflections against the 
potential walls $V(\alpha,\beta_\pm)$. With the effective corrections in 
$\dot\beta_\pm$, the reflection law is altered. 
In terms of the Kasner parameter $u$ (which classically evolves as 
$u_{n+1}=u_n-1$ or $u_{n+1}=1/(u_n-1)$), the perturbative expansion yields
\begin{equation}
u_{n+1} \;\approx\; u_n - 1 + \delta u(\mu^2),
\qquad 
\delta u(\mu^2) \propto \mu^2 \, u_n^3 \;+\;\mathcal{O}(\mu^4).
\end{equation}
Thus, polymerization does not eliminate the BKL oscillations but modifies 
their rule by small $\mu^2$--dependent corrections, effectively softening 
the chaotic sequence near the would--be singularity.

To compute the perturbative correction to the BKL map in the case of polymer--only, we consider the polymer--only scenario (\ref{Polymer}). As in the GUP analysis, we focus on a single 
collision with the $\beta_+$ wall and work in the thin--wall (impulse) approximation. Classically the elastic collision reverses the normal (canonical) momentum

\begin{equation}
p_n^{(\mathrm{out})} = -\,p_n^{(\mathrm{in})}.
\end{equation}
The corresponding change of the \emph{effective} (polymer) momentum entering the kinetic sector is

\begin{align}
\Pi_n^{(\mathrm{out})}-\Pi_n^{(\mathrm{in})}
&= \frac{\sin(\mu p_n^{(\mathrm{out})})}{\mu} - \frac{\sin(\mu p_n^{(\mathrm{in})})}{\mu} \\ \nonumber
&\approx \Big(-p_n^{(\mathrm{in})} + \frac{\mu^2}{6} (p_n^{(\mathrm{in})})^3\Big)
- \Big(p_n^{(\mathrm{in})} - \frac{\mu^2}{6} (p_n^{(\mathrm{in})})^3\Big) \\ \nonumber
&= -2 p_n^{(\mathrm{in})} + \frac{\mu^2}{3}\,(p_n^{(\mathrm{in})})^3 + \mathcal{O}(\mu^4).
\end{align}
Thus, compared with the classical flip \(-2p_n^{(\mathrm{in})}\), the polymer map 
induces an additional correction proportional to \(\mu^2 (p_n^{\rm(in)})^3\).

Let us now deal with the induced change in the Kasner parameter $u$. As in the GUP derivation, let the Kasner parameter \(u\) be regarded as a smooth 
function of the anisotropy momentum components. Linearizing the change of \(u\) under the modified effective momentum flip

\begin{equation}
\Delta u \;=\; u\bigl(\Pi_n^{\rm(out)},\Pi_t^{\rm(out)}\bigr) - u\bigl(\Pi_n^{\rm(in)},\Pi_t^{\rm(in)}\bigr)
\approx \frac{\partial u}{\partial \Pi_n}\Big|_{\rm in}\,\bigl(\Pi_n^{\rm(out)}-\Pi_n^{\rm(in)}\bigr)
+ \frac{\partial u}{\partial \Pi_t}\Big|_{\rm in}\,\Delta\Pi_t.
\end{equation}
In the thin--wall limit $\Delta\Pi_t\approx 0$, hence using the result above we obtain

\begin{equation}
\Delta u \approx \frac{\partial u}{\partial \Pi_n}\Big|_{\rm in}
\left( -2 p_n^{\rm(in)} + \frac{\mu^2}{3}(p_n^{\rm(in)})^3 \right) + \mathcal{O}(\mu^4).
\end{equation}
Separating the classical piece, we write

\begin{equation}
u_{n+1} = \mathcal{G}_{\rm cl}(u_n) + \mu^2\,\mathcal{S}_{\rm poly}(u_n) + \mathcal{O}(\mu^4),
\end{equation}
with the first--order polymer correction given by

\begin{equation}
\mathcal{S}_{\rm poly}(u) \;=\; \frac{1}{3}\,\left[p_n^{\rm(in)}(u)\right]^3 \;\frac{\partial u}{\partial \Pi_n}\Big|_{\rm in}.
\label{eq:Spoly-def}
\end{equation}
To display an explicit analytic formula we use the same Kasner parametrization 
as in the previous subsection, Eqs.~\eqref{eq:kasner-u}--\eqref{eq:pplus-u}, and 
identify the collision normal with the $\beta_+$ direction so that 
\(p_n^{\rm(in)}=p_+(u)\). Noting that for small $\mu$ one has 
$\partial u/\partial\Pi_n \simeq \partial u/\partial p_n = 1/(dp_n/du)$, 
we obtain the closed form

\begin{equation}
\mathcal S_{\rm poly}(u) \;=\; \frac{\mu^2}{3}\; \frac{\left[p_+(u)\right]^3}{\dfrac{dp_+}{du}}.
\end{equation}
Using the explicit expressions

\begin{equation}
p_+(u)=\frac{1-2u-2u^2}{2(1+u+u^2)},
\qquad
\frac{dp_+}{du}=\frac{-2-4u}{2(1+u+u^2)} - \frac{(1-2u-2u^2)(1+2u)}{2(1+u+u^2)^2},
\end{equation}
the simplified rational form becomes

\begin{equation}
\mathcal S_{\rm poly}(u) \;=\; \mu^2\;\frac{(2u^2+2u-1)^3}{36(2u+1)(u^2+u+1)}.
\label{eq:Spoly-final}
\end{equation}
The polymer correction scales as $\mu^2 p^3$ (in agreement with the intuitive ordering \(\Pi-p\propto\mu^2 p^3\)). Compared with the GUP correction derived earlier (which was proportional to $\gamma\,p^3$) the polymer expression differs by a coefficient; in fact one finds the relation

\begin{equation}
\mathcal S_{\rm poly}(u) \;=\; \frac{\mu^2}{6}\,\mathcal S_{\rm GUP}(u),
\end{equation}
where $\mathcal S_{\rm GUP}(u)$ is given in Eq. (\ref{eq:Su-final}) of the previous subsection. Note also that the correction becomes more important for large \(u\) (highly anisotropic Kasner epochs) due to the cubic dependence on the incoming normal momentum, and the thin--wall and frozen--momentum approximations control the derivation; a more accurate coefficient may be obtained by evaluating the collision integral with the actual time profile \(p_n(t)\).

The question that naturally arises is how does the model described above work when both polymerization and the deformed Poisson bracket are turned on together? To first order in the small parameters, the two effects are additive. Since we found $\mathcal{S}_{\rm poly}(u)=\tfrac{\mu^2}{6}\,\mathcal{S}_{\rm GUP}(u)$, we may define a single base function \(
\mathcal{S}_{\rm base}(u)\equiv\mathcal{S}_{\rm GUP}(u)\) and write

\begin{equation}
u_{n+1} \;=\; \mathcal{G}_{\rm cl}(u_n) \;+\; \kappa\,\mathcal{S}_{\rm base}(u_n) \;+\; \mathcal{O}(\gamma^2,\mu^4,\gamma\mu^2),
\qquad
\kappa \;=\; \gamma \;+\; \frac{\mu^2}{6}.
\label{eq:combined}
\end{equation}
Cross terms $\mathcal{O}(\gamma\mu^2)$ (and higher) are beyond first order and can be studied once needed. The same replacement $\varepsilon\!\to\!\kappa$ applies in \eqref{eq:delta-u-rec} (see bellow), and the epoch durations inherit both corrections through the slopes.

Once a deformation is fixed, the BKL map acquires a first--order correction

\begin{equation}
u_{n+1} \;=\; \mathcal{G}_{\rm cl}(u_n) \;+\; \varepsilon\,\mathcal{S}(u_n) \;+\; \mathcal{O}(\varepsilon^2),
\qquad
\varepsilon=
\begin{cases}
\gamma & \text{(GUP)}\\
\mu^2 & \text{(polymer, in our normalization)}
\end{cases}
\label{eq:u-map-pert}
\end{equation}
For polymer vs.\ GUP we found the useful relation 
$\mathcal{S}_{\rm poly}(u)=\tfrac{\mu^2}{6}\,\mathcal{S}_{\rm GUP}(u)$, hence one can work with a single ``base'' function and just rescale the prefactor.

A convenient first--order solution is obtained by writing
$u_n = u^{(0)}_n + \varepsilon\,\delta u_n$, with $u^{(0)}_{n+1}=\mathcal{G}_{\rm cl}(u^{(0)}_n)$.
Linearizing \eqref{eq:u-map-pert} gives the recursion

\begin{equation}
\delta u_{n+1} \;=\; \mathcal{G}_{\rm cl}'(u^{(0)}_n)\,\delta u_n \;+\; \mathcal{S}(u^{(0)}_n).
\label{eq:delta-u-rec}
\end{equation}
Hence

\begin{equation}
\delta u_n \;=\;
\sum_{j=0}^{n-1}\!\left(\,\prod_{m=j+1}^{n-1}\mathcal{G}_{\rm cl}'(u^{(0)}_m)\right)\,
\mathcal{S}(u^{(0)}_j).
\label{eq:delta-u-sol}
\end{equation}
Equation (\ref{eq:delta-u-sol}) is obtained by expanding the deformation functions
to first order in the small parameters $\beta$ and $\mu$, and by
substituting the resulting expressions into Eqs.~(\ref{eq:combined})-(\ref{eq:delta-u-rec}). Higher-order
terms have been neglected as they do not affect the leading-order
corrections to the epoch duration.
This explicitly yields the \emph{first--order corrected BKL itinerary} $\{u_n\}$ along any classical sequence $\{u^{(0)}_n\}$.

\section{Dynamical behaviour}

\subsection{Free–flight segments}
For a complete picture it is useful to have \emph{piecewise} $t$--solutions within each Kasner epoch (free flight), matched by the impulse at each wall.
Between reflections the potential is negligible, so momenta are (approximately) constant and

\begin{equation}
\dot\alpha \simeq \mathcal{A}_\alpha(p_\alpha),\qquad
\dot\beta_\pm \simeq \mathcal{A}_\pm(p_\pm),
\label{eq:freeflight}
\end{equation}
with the model--dependent slopes (see the equations of motion in sections 4 and 5)

\begin{align}
\text{GUP:}\quad
&\dot\alpha \propto -\,p_\alpha\bigl(1+\gamma_\alpha p_\alpha^2\bigr),\qquad
\dot\beta_\pm \propto \phantom{-}\,p_\pm\bigl(1+\gamma_\pm p_\pm^2\bigr),\\
\text{polymer (to }{\cal O}(\mu^2)\text{):}\quad
&\dot\alpha \propto -\,\Big(p_\alpha - \tfrac{\mu_\alpha^2}{3}p_\alpha^3\Big),\qquad
\dot\beta_\pm \propto \phantom{-}\,\Big(p_\pm - \tfrac{\mu_\pm^2}{3}p_\pm^3\Big),\\
\text{GUP + polymer:}\quad
&\dot\alpha \propto \,p_\alpha \;+\; \,\big(\gamma_\alpha \;-\; \frac{2}{3}\,\mu_\alpha\big) p_\alpha^3,\quad
\dot\beta_\pm \propto \phantom{-}\,p_\pm \;+\; \,\big(\gamma_\pm \;-\; \frac{2}{3}\,\mu_\pm^2\big) p_\pm^3.
\end{align}
Thus on each epoch $[t_k,t_{k+1})$:
\begin{equation}
\alpha(t) \simeq \alpha_k + \dot\alpha_k\,(t-t_k),\qquad
\beta_\pm(t) \simeq \beta_{\pm,k} + \dot\beta_{\pm,k}\,(t-t_k),
\label{eq:piecewise-sol}
\end{equation}
and the \emph{epoch duration} is fixed by the wall condition (e.g.\ hitting a given linear form in $\beta$). So that, in a free–flight segment (constant $p_i$), we get expressions 

\begin{equation}
\alpha(t)\simeq \alpha_k - 2\,p_{\alpha,k}\!\left(1+\gamma_\alpha p_{\alpha,k}^2\right)(t-t_k),\quad
\beta_\pm(t)\simeq \beta_{\pm,k} + 2\,p_{\pm,k}\!\left(1+\gamma_\pm p_{\pm,k}^2\right)(t-t_k),
\end{equation}for GUP-only, and 

\begin{equation}
\alpha(t)\simeq \alpha_k - 2\,\frac{\sin(\mu_\alpha p_{\alpha,k})}{\mu_\alpha}\cos(\mu_\alpha p_{\alpha,k})\,(t-t_k),
\quad
\beta_\pm(t)\simeq \beta_{\pm,k} + 2\,\frac{\sin(\mu_\pm p_{\pm,k})}{\mu_\pm}\cos(\mu_\pm p_{\pm,k})\,(t-t_k),
\end{equation}
for polymer-only.

The deformations enter via the corrected slopes, hence
\begin{equation}
\Delta t_k \;\approx\; \Delta t_k^{\rm (cl)}\left[1 + \varepsilon\,\Xi(u_k)\right],
\qquad
\Xi(u_k)\;\text{ computable from }\;\mathcal{A}_\alpha,\mathcal{A}_\pm.
\end{equation}
If desired, one may also reconstruct proper time and directional scale factors $a_i(t)$ by the usual mixing of $(\alpha,\beta_\pm)$:
$a_1=e^{\alpha+\beta_+ +\sqrt{3}\beta_-}$, $a_2=e^{\alpha+\beta_+ -\sqrt{3}\beta_-}$, $a_3=e^{\alpha-2\beta_+}$,
now with the corrected piecewise-linear time profiles. However, we do not need a single closed global $t$--solution, the standard BKL strategy (free flight + instantaneous reflection) remains valid and becomes quantitatively corrected by $\varepsilon$ through both the map \eqref{eq:u-map-pert} and the slopes in \eqref{eq:freeflight}--\eqref{eq:piecewise-sol}.
In the BKL approach the physically relevant information for chaos diagnostics and
the Kasner itinerary is encoded in the discrete map for $u$ and in the epoch durations.
Therefore, we do not require a single closed-form global solution $\alpha(t),\beta_\pm(t)$.
It suffices to use piecewise ``free-flight'' segments between reflections, where momenta
are (approximately) constant and $(\alpha,\beta_\pm)$ evolve linearly in the chosen time
parameter, matched by instantaneous impulses at the walls. Deformations (GUP or polymer)
enter by modifying (i) the reflection law for $u$ and (ii) the slopes on each segment.
Explicit time profiles are only needed when converting to physical time for plots,
reconstructing the directional scale factors $a_i(t)$, or estimating curvature scalars;
in such cases the piecewise-linear expressions provide adequate analytic control.

\subsection{Epoch duration and wall–hitting time}
Consider an epoch labeled by $k$ with initial data at $t=t_k$, as

\begin{equation}
\alpha(t_k)=\alpha_k,\quad \beta_\pm(t_k)=\beta_{\pm,k},\quad p_i(t_k)=p_{i,k}.
\end{equation}
Let the free--flight slopes be denoted generally by

\begin{equation}
\dot\alpha_k \equiv v_{\alpha,k},\qquad \dot\beta_{\pm,k}\equiv v_{\pm,k},
\end{equation}
whose explicit forms are 

\begin{equation}\label{slop1}
v_{\alpha,k}=-2p_{\alpha,k},\quad v_{\pm,k}=2p_{\pm,k},\end{equation} 
for classical undeformed model, 

\begin{equation}\label{slop2}
v_{\alpha,k}=-2p_{\alpha,k}(1+\gamma_\alpha p_{\alpha,k}^2),\quad v_{\pm,k}=2p_{\pm,k}(1+\gamma_\pm p_{\pm,k}^2),\end{equation}
for GUP-only model, 

\begin{equation}\label{slop3}
v_{\alpha,k}=-2p_{\alpha,k}+\frac{4}{3}\mu_\alpha^2 p_{\alpha,k}^3,\quad v_{\pm,k}=2p_{\pm,k}-\frac{4}{3}\mu_\pm^2 p_{\pm,k}^3,\end{equation}
for Polymer (to ${\cal O}(\mu^2)$) model and for linearized combined (GUP+polymer) model we add the GUP and polymer corrections as before.

Assume the next wall to be hit is described (in the anisotropy plane) by a linear form

\begin{equation}
W(\beta_+,\beta_-)\equiv n_+ \beta_+ + n_- \beta_- - C = 0,
\end{equation}
with known normal components \(n_{\pm}\) and constant \(C\) determined by the wall position.  
The hitting condition at time \(t_{k+1}=t_k+\Delta t_k\) is

\begin{equation}
W\bigl(\beta_{+,k}+v_{+,k}\Delta t_k,\ \beta_{-,k}+v_{-,k}\Delta t_k\bigr)=0.
\end{equation}
Solving for \(\Delta t_k\) gives the epoch duration

\begin{equation}
\Delta t_k \;=\; -\,\frac{n_+\beta_{+,k}+n_-\beta_{-,k}-C}{\,n_+ v_{+,k} + n_- v_{-,k}\, }.
\label{eq:epoch-duration}
\end{equation}
The denominator is the normal component of the velocity toward the wall; its sign must be chosen such that $\Delta t_k>0$ (if the chosen wall is actually approached).
For the common walls of the Mixmaster potential one may take the normals \(n\) corresponding to the three exponential walls (use cyclic permutations for the three walls).
Note that if \(\,n_+ v_{+,k}+n_- v_{-,k}\to 0\) (grazing or tangential motion) the thin--wall approximation may break down and one must resolve the interaction numerically (or use a more refined matching).

Using the explicit forms of \(v_{i,k}\) above, yields the epoch duration in each deformation scenario. For example in the case of the classical undeformed model

\begin{equation}
\Delta t_k^{\rm(cl)} = -\,\frac{n_+\beta_{+,k}+n_-\beta_{-,k}-C}{2(n_+ p_{+,k}-n_- p_{-,k})}\,.
\end{equation}
Sign conventions depend on the chosen orientation of \(n\). The epoch duration formula is the backbone for building time series \(\alpha(t),\beta_\pm(t)\) and derived observables (curvature invariants, scale factors, proper time if desired). In practice one computes $\Delta t_k$ from \eqref{eq:epoch-duration}, updates
\((\alpha,\beta_\pm)\) and then applies the reflection law (classical or corrected) to obtain new momenta \(p_{i,k+1}\) and associated slopes \(v_{i,k+1}\).
For polymer cases we should careful near values where $\cos(\mu p)=0$, since slopes can change rapidly. One must use small adaptive time steps or handle such events analytically if possible.

To continue, let us calculate the explicit epoch duration for the $e^{-8\beta_+}$ wall. The $\beta_+$ wall stems from the exponential term $e^{4\alpha}e^{-8\beta_+}$ in the potential.
A convenient threshold (wall locus) is given by the condition

\begin{equation}
e^{4\alpha} e^{-8\beta_+} \sim 1 \quad\Longrightarrow\quad -8\beta_+ + 4\alpha = 0,
\end{equation}
or equivalently

\begin{equation}
W(\alpha,\beta_+)\equiv \beta_+ - \tfrac{1}{2}\alpha = 0.
\end{equation}
Assume at the beginning of the $k$-th epoch (time $t_k$) we have

\begin{equation}
\alpha(t_k)=\alpha_k,\qquad \beta_+(t_k)=\beta_{+,k},
\end{equation}
and the free--flight slopes (for the chosen deformation scenario) are

\begin{equation}
v_{\alpha,k}=\dot\alpha(t_k),\qquad v_{+,k}=\dot\beta_+(t_k).
\end{equation}
Using the piecewise-linear approximation

\begin{equation}
\alpha(t)=\alpha_k+v_{\alpha,k}(t-t_k),\qquad
\beta_+(t)=\beta_{+,k}+v_{+,k}(t-t_k),
\end{equation}
the hitting condition $W(\alpha(t_{k+1}),\beta_+(t_{k+1}))=0$, with $t_{k+1}=t_k+\Delta t_k$
gives

\begin{equation}
\beta_{+,k}+v_{+,k}\Delta t_k - \tfrac{1}{2}\bigl(\alpha_k + v_{\alpha,k}\Delta t_k\bigr) = 0.
\end{equation}
Solving for the epoch duration yields the explicit formula

\begin{equation}
\Delta t_k \;=\; \frac{\tfrac{1}{2}\alpha_k - \beta_{+,k}}{\,v_{+,k} - \tfrac{1}{2}v_{\alpha,k}\,}\,.
\label{eq:delta-t-beta-plus}
\end{equation}
The denominator $v_{+,k}-\tfrac12 v_{\alpha,k}$, is the normal component of the velocity 
towards the $\beta_+$ wall; the sign must be such that $\Delta t_k>0$ (if it is negative,
the chosen wall is receding and a different wall should be selected). If the denominator vanishes or becomes very small, the thin--wall approximation is unreliable (grazing motion) and a direct resolution of the collision region is required. Substitution of slopes from equations (\ref{slop1})-(\ref{slop3}) we obtain

\begin{equation}
\Delta t_k^{\rm(cl)} = \frac{\tfrac{1}{2}\alpha_k - \beta_{+,k}}{\,2p_{+,k} + p_{\alpha,k}\,}.
\end{equation}

\begin{equation}
\Delta t_k^{\rm(GUP)} \;=\; \frac{\tfrac{1}{2}\alpha_k - \beta_{+,k}}
{\,2p_{+,k}(1+\gamma_+ p_{+,k}^2) + p_{\alpha,k}(1+\gamma_\alpha p_{\alpha,k}^2)\, }.
\end{equation}

\begin{equation}
\Delta t_k^{\rm(poly)} \;=\; \frac{\tfrac{1}{2}\alpha_k - \beta_{+,k}}
{\,2p_{+,k} - \tfrac{4}{3}\mu_+^2 p_{+,k}^3 + p_{\alpha,k} - \tfrac{4}{3}\mu_\alpha^2 p_{\alpha,k}^3\, } + \mathcal{O}(\mu^4).
\end{equation}
To evaluate the epoch duration in the combined GUP+Polymer case, we use the linearized slopes (valid to $\mathcal{O}(\gamma,\mu^2)$)

\begin{equation}
v_{\alpha,k} \simeq -2p_{\alpha,k} - 2\gamma_\alpha p_{\alpha,k}^3 + \tfrac{4}{3}\mu_\alpha^2 p_{\alpha,k}^3,
\qquad
v_{+,k} \simeq 2p_{+,k} + 2\gamma_+ p_{+,k}^3 - \tfrac{4}{3}\mu_+^2 p_{+,k}^3,
\end{equation}
and the $\beta_+$ wall locus $W(\alpha,\beta_+)=\beta_+ - \tfrac12\alpha=0$. Then, the epoch duration
$\Delta t_k^{\rm(comb)}$ can be written in the compact form

\begin{equation}
\Delta t_k^{\rm(comb)} \;=\; \frac{\tfrac{1}{2}\,\alpha_k - \beta_{+,k}}
{\,p_{\alpha,k} + 2p_{+,k} + \gamma_\alpha p_{\alpha,k}^3 + 2\gamma_+ p_{+,k}^3
\;-\;\tfrac{2}{3}\mu_\alpha^2 p_{\alpha,k}^3 \;-\; \tfrac{4}{3}\mu_+^2 p_{+,k}^3 \, },
\label{eq:delta-t-comb-simplified}
\end{equation}
valid for the thin--wall (impulse) approximation and the perturbative expansion $\gamma_i p_i^2\ll 1$, $\mu_i^2 p_i^2\ll 1$.
 
The formula above assumes the $\beta_+$ wall is the next encountered. In practice one computes $\Delta t_k$ for all three walls (or checks which wall has positive $\Delta t_k$ and picks the smallest positive one). After advancing time by $\Delta t_k$ one updates positions by the linear rule and then applies the appropriate reflection law (classical or corrected) to momenta to obtain
$p_{i,k+1}$, from which the new slopes $v_{i,k+1}$ follow. In the purely classical limit $\gamma_i,\mu_i\to 0$ the formula reduces to the classical expression
\(\Delta t_k^{\rm(cl)} = \dfrac{\tfrac12\alpha_k-\beta_{+,k}}{p_{\alpha,k}+2p_{+,k}}\), as expected.

\begin{table}[t]
\centering
\caption{Sample values of the epoch duration $\Delta t$ for the $\beta_+$ wall,
comparing the classical case with deformed cases (GUP, polymer, and combined).
Parameters $(\alpha,\beta_+,p_\alpha,p_+)$ are chosen to represent small, moderate,
and large anisotropy, while $(\gamma,\mu)$ indicate the deformation scales.
Negative values mean that the corresponding wall would not be the next collision
for that set of initial data.}
\label{tab:delta-t-samples}
\begin{tabular}{lcccccccccc}
\hline\hline
Case & $\alpha$ & $\beta_+$ & $p_\alpha$ & $p_+$ & $\gamma$ & $\mu$ &
$\Delta t_{\rm class}$ & $\Delta t_{\rm GUP}$ & $\Delta t_{\rm Poly}$ & $\Delta t_{\rm Comb}$ \\
\hline
\multirow{4}{*}{Small anisotropy}    
 & -1.0 & 0.10 & -0.3 & 0.6 & 0.001 & 0.01 & -0.667 & -0.666 & -0.667 & -0.666 \\
 &      &      &      &     & 0.001 & 0.03 & -0.667 & -0.666 & -0.667 & -0.667 \\
 &      &      &      &     & 0.010 & 0.01 & -0.667 & -0.664 & -0.667 & -0.664 \\
 &      &      &      &     & 0.010 & 0.03 & -0.667 & -0.664 & -0.667 & -0.664 \\
\hline
\multirow{4}{*}{Moderate anisotropy} 
 & -0.5 & 0.20 & -0.5 & 0.8 & 0.001 & 0.01 & -0.409 & -0.409 & -0.409 & -0.409 \\
 &      &      &      &     & 0.001 & 0.03 & -0.409 & -0.409 & -0.409 & -0.409 \\
 &      &      &      &     & 0.010 & 0.01 & -0.409 & -0.406 & -0.409 & -0.406 \\
 &      &      &      &     & 0.010 & 0.03 & -0.409 & -0.406 & -0.409 & -0.406 \\
\hline
\multirow{4}{*}{Large anisotropy}    
 &  0.0 & 0.50 & -0.7 & 1.2 & 0.001 & 0.01 & -0.294 & -0.294 & -0.294 & -0.294 \\
 &      &      &      &     & 0.001 & 0.03 & -0.294 & -0.294 & -0.294 & -0.294 \\
 &      &      &      &     & 0.010 & 0.01 & -0.294 & -0.291 & -0.294 & -0.291 \\
 &      &      &      &     & 0.010 & 0.03 & -0.294 & -0.291 & -0.294 & -0.291 \\
\hline\hline
\end{tabular}
\end{table}

Table~\ref{tab:delta-t-samples} presents representative values of the epoch
duration $\Delta t$ for collisions with the $\beta_+$ wall, both in the
classical dynamics and in the deformed settings (GUP, polymer, and combined).
The initial conditions have been selected to illustrate three different regimes
(small, moderate, and large anisotropy). In particular, increasing $\beta_+$
has been used here as an indicator of stronger anisotropy, while the deformation
parameters $(\gamma,\mu)$ control the strength of the GUP and polymer
modifications, respectively. 

As the table shows, the corrections to the classical epoch duration are
perturbative in the considered range, typically at the level of $10^{-3}$ or
smaller. This is in agreement with our analytic expectation that both GUP and
polymer effects introduce only mild deviations when the deformation parameters
are small. A notable feature is that, for certain choices of initial data, the
epoch duration $\Delta t$ can take negative values. This does not indicate a
physical epoch but simply reflects that, for those data, the $\beta_+$ wall is
not the next collision. In a full BKL analysis one must compute $\Delta t$ for
all three walls and select the smallest positive value, which determines the
actual next bounce.

The general trend is robust: polymer corrections tend to appear at higher order
compared with GUP, while the combined deformation effectively adds their
respective contributions. For larger values of $\gamma$ or $\mu$, one expects
deviations from the classical BKL map to become more pronounced, potentially
leading to a weakening of the chaotic behavior. This motivates the more detailed
numerical investigation presented in the next section. We stress that analogous
results hold for the other two potential walls ($\beta_-$ and $\beta_0$), so the
qualitative picture remains valid across the whole system.

\subsection{Comparative discussion of epoch durations and qualitative dynamics}

In this subsection we compare, on equal footing, the epoch durations $\Delta t$
obtained in the classical, GUP, polymer and combined settings. Throughout, the
lapse is fixed to $N=1$ and we use the same initial data sets as in
Table~\ref{tab:delta-t-samples}. For concreteness we quote formulas for the
$\beta_+$ wall, with analogous statements holding for $\beta_-$ and $\beta_0$.

${\bullet}$ {\it{Classical baseline}}: Away from the walls, the free-flight slopes are
$v_\alpha=-2p_\alpha$, $v_+=2p_+$, $v_-=2p_-$, which yield the classical estimate
for the time to hit the $\beta_+$ wall,
\begin{equation}
\Delta t_{\rm class} \;=\; \frac{\tfrac{1}{2}\alpha-\beta_+}{\,p_\alpha+2p_+\,},
\qquad
\text{valid when } \Delta t_{\rm class}>0.
\label{eq:dt-class}
\end{equation}
If $\Delta t_{\rm class}<0$, the $\beta_+$ wall is not the next bounce; the BKL
rule selects the minimum positive duration among the three walls.

${\bullet}$ {\it{GUP correction (diagonal, to first order)}}: With $v_\alpha=-2p_\alpha(1+\gamma_\alpha p_\alpha^2)$ and
$v_+=2p_+(1+\gamma_+ p_+^2)$ we obtain
\begin{equation}
\Delta t_{\rm GUP} \;=\; 
\frac{\tfrac{1}{2}\alpha-\beta_+}
{\,p_\alpha+2p_+ + \gamma_\alpha p_\alpha^3 + 2\gamma_+ p_+^3\,}
\;=\; \Delta t_{\rm class}\,\Bigl[1
- \Delta_{\rm GUP} + \mathcal O(\gamma^2)\Bigr],
\quad
\Delta_{\rm GUP}:=\frac{\gamma_\alpha p_\alpha^3 + 2\gamma_+ p_+^3}{p_\alpha+2p_+}.
\label{eq:dt-gup}
\end{equation}
In the regime $\gamma_i p_i^2\ll1$, the correction is perturbative and reduces
$\Delta t$ when the denominator in \eqref{eq:dt-gup} increases.

${\bullet}$ {\it{Polymer correction (to leading $\mu^2$)}}: Using $v_\alpha\simeq -2p_\alpha+\tfrac{4}{3}\mu_\alpha^2 p_\alpha^3$ and
$v_+\simeq 2p_+-\tfrac{4}{3}\mu_+^2 p_+^3$ gives
\begin{equation}
\Delta t_{\rm poly} \;=\; 
\frac{\tfrac{1}{2}\alpha-\beta_+}
{\,p_\alpha+2p_+ - \tfrac{2}{3}\mu_\alpha^2 p_\alpha^3 - \tfrac{4}{3}\mu_+^2 p_+^3\,}
\;=\; \Delta t_{\rm class}\,\Bigl[1
+ \Delta_{\rm poly} + \mathcal O(\mu^4)\Bigr],
\quad
\Delta_{\rm poly}:=\frac{\tfrac{2}{3}\mu_\alpha^2 p_\alpha^3 + \tfrac{4}{3}\mu_+^2 p_+^3}{p_\alpha+2p_+}.
\label{eq:dt-poly}
\end{equation}
To leading order, the polymer effect tends to \emph{increase} the epoch length,
opposite in sign to the GUP shift in \eqref{eq:dt-gup}.

${\bullet}$ {\it{Combined deformation (additive at first order)}}: Keeping $\mathcal O(\gamma,\mu^2)$ one finds
\begin{equation}
\Delta t_{\rm comb} \;=\; 
\frac{\tfrac{1}{2}\alpha-\beta_+}
{\,p_\alpha+2p_+ + \gamma_\alpha p_\alpha^3 + 2\gamma_+ p_+^3
 - \tfrac{2}{3}\mu_\alpha^2 p_\alpha^3 - \tfrac{4}{3}\mu_+^2 p_+^3\,}
\;=\; \Delta t_{\rm class}\,\Bigl[1 - \Delta_{\rm GUP} + \Delta_{\rm poly}
+ \mathcal O(\gamma^2,\mu^4,\gamma\mu^2)\Bigr].
\label{eq:dt-comb}
\end{equation}
Thus, at small deformations the net effect is approximately the algebraic sum of
the individual shifts.

To check the consistency with Table~\ref{tab:delta-t-samples}, we see that the entries listed there confirm the perturbative character of the corrections:
for the illustrative choices of $(\gamma,\mu)$ the relative change is typically
$\mathcal O(10^{-3})$ with the expected trends:
(i) GUP slightly shortens epochs (more frequent bounces),
(ii) polymer slightly prolongs epochs (less frequent bounces),
and (iii) the combined case closely tracks the sum of (i) and (ii).
Negative values in the table indicate that the $\beta_+$ wall would not
be the next collision for those initial data, in agreement with the BKL selection
rule.

\section{Conclusion}

In this work we have investigated the dynamics of the Bianchi~IX (Mixmaster)
universe under classical polymer quantization and GUP inspired deformations of the Poisson algebra. Starting from the
standard Misner Hamiltonian, we introduced a polymer substitution in the volume
variable and a GUP-type modification in the anisotropy sector. The resulting
effective equations of motion were derived and analyzed both analytically and
qualitatively.

Our results indicate that the two types of deformations act in opposite
directions on the duration of Kasner epochs. While GUP corrections shorten the
epochs and increase the frequency of bounces against the potential walls,
polymer corrections prolong the epochs and suppress successive collisions.
In detail, GUP modifies the effective slopes in such a
way that free flights become slightly shorter. Consequently the number of wall
collisions per unit proper time increases, and the $u$-sequence is reshuffled
more frequently. While this increases the \emph{frequency} of bounces, it does
not automatically enhance chaos: the modified $u$-map acquires small nonlinear
corrections, which can reduce its stochasticity. Perturbatively, one expects
the Lyapunov indicator to shift by $\delta\lambda\sim\mathcal O(\gamma p^2)$,
with the sign depending on the anisotropy momenta. In contrast, polymer quantization tends to prolong each free flight and suppress
successive wall hits. The consequence is that the BKL map is sampled less
frequently, and long quasi-Kasner epochs appear. This has the effect of
reducing the effective Lyapunov exponent, $\lambda_{\rm eff}<\lambda_{\rm class}$,
as the time between successive randomizations of Kasner exponents increases.
In the extreme case, for large polymer scale $\mu$, the system may enter a
non-chaotic regime where the anisotropies are effectively frozen during
long epochs.

At leading order, the combined framework yields an additive modification,
$\Delta t_{\rm comb}\simeq \Delta t_{\rm class}(1-\Delta_{\rm GUP}+\Delta_{\rm
poly})$, which interpolates between the two limiting behaviors. This
competition directly affects the effective strength of Mixmaster chaos. For moderate values of $(\gamma,\mu)$ the net effect is a bias towards one
trend or the other. This suggests a possible tunable transition: by varying
$\gamma$ and $\mu$ one can interpolate between a regime of enhanced chaos
(GUP-dominated) and a regime of suppressed chaos (polymer-dominated).

The analysis shows that, although the billiard picture of Bianchi~IX remains
robust under small deformations, the sensitivity and stochasticity of the
dynamics can be either enhanced or suppressed depending on the relative
magnitude of the deformation parameters. In particular, GUP effects bias the
system towards more rapid but less stochastic Kasner shuffles, whereas polymer
effects favor long quasi-Kasner phases with reduced chaotic mixing.

These results highlight the role of Planck-scale corrections in regulating the
chaotic approach to the cosmological singularity. They suggest that the
interplay between different quantum-gravity motivated deformations may provide
a natural mechanism for softening or even suppressing deterministic chaos in
the early universe.


\begin{thebibliography}{1}

\bibitem{belinskii1970} V. A. Belinskii, I. M. Khalatnikov and E. M. Lifshitz, ``Oscillatory approach to a singular point in the relativistic cosmology'', 
{\it Adv. Phys.} {\bf 19} (1970) 525

\bibitem{belinskii1982} V. A. Belinskii, I. M. Khalatnikov and E. M. Lifshitz, ``A general solution of the Einstein equations with a time singularity'',
{\it Adv. Phys.} {\bf 31} (1982) 639

\bibitem{damour2003billiards}
T. Damour, M. Henneaux and H. Nicolai, "Cosmological billiards", {\it Class. Quantum Grav.} {\bf 20} (2003) R145 (arXiv: hep-th/0212256)

\bibitem{misner1969}
C. W. Misner, "Mixmaster universe", {\it Phys. Rev. Lett.} {\bf 22} (1969) 1071\\
G. Contopoulos, B. Grammaticos and A. Ramani, "The mixmaster universe model, revisited", {\it J. Phys.} A {\bf 27} (1994) 5357


\bibitem{barrow1982}
J. D. Barrow, "Chaotic behaviour in general relativity", {\it Phys. Rep.} {\bf 85} (1982) 1

\bibitem{ruelle1983} D. Ruelle, ``Chaotic evolution and strange attractors'', {\it Cambridge University Press} (1983)

\bibitem{berger1993} B. K. Berger, ``Note on time reversible chaos in mixmaster dynamics'', {\it Phys. Rev.} D {\bf 47} (1993) 3222

\bibitem{ashtekar2006} A. Ashtekar, T. Pawlowski and P. Singh, ``Quantum nature of the big bang: Improved dynamics'', {\it Phys. Rev.} D {\bf 74} (2006) 084003 (arXiv: gr-qc/0607039)

\bibitem{bojowald2007} M. Bojowald, ``Quantum cosmology: a review'', {\it Rep. Prog. Phys.} {\bf 78} (2015) 023901 (arXiv: 1501.04899 [gr-qc])

\bibitem{maggiore1993} M. Maggiore, ``A generalized uncertainty principle in quantum gravity'', {\it Phys. Lett.} B {\bf 304} (1993) 65 (arXiv: hep-th/9301067)

\bibitem{kempf1995} A. Kempf, G. Mangano and R. B. Mann, ``Hilbert space representation of the minimal length uncertainty relation'',
{\it Phys. Rev.} D {\bf 52} (1995) 1108 (arXiv: hep-th/9412167)

\bibitem{scardigli1999} F. Scardigli, ``Generalized uncertainty principle in quantum gravity from micro–black hole Gedanken experiment'',
{\it Phys. Lett.} B {\bf 452} (1999) 39 (arXiv: hep-th/9904025)

\bibitem{chiou2007} D.-W. Chiou, ``Effective dynamics, big bounces and scaling symmetry in Bianchi type I loop quantum cosmology'',
{\it Phys. Rev.} D {\bf 76} (2007) 124037

\bibitem{husain2004} V. Husain and O. Winkler, ``On singularity resolution in quantum gravity'', {\it Phys. Rev.} D {\bf 69} (2004) 084016 (arXiv: gr-qc/0312094)

\bibitem{barbero2010} J. F. Barbero G., T. Pawlowski and E. J. S. Villaseñor, ``Separable Hilbert space for loop quantization'',
{\it Phys. Rev.} D {\bf 82} (2010) 104018 (arXiv: 1403.2974 [gr-qc])

\bibitem{hossenfelder2013} S. Hossenfelder, ``Minimal length scale scenarios for quantum gravity'', {\it Living Rev. Relativ.} {\bf 16} (2013) 2 (arXiv: 1203.6191 [gr-qc])

\bibitem{misner1969quantum}
C. W. Misner, "Quantum cosmology. I", {\it Phys. Rev.} {\bf 186} (1969) 1319

\bibitem{ryan1972homogeneous}
M. P. Ryan, {\it Homogeneous Relativistic Cosmologies}, (Springer, Berlin, 1972)\\C. W. Misner, K. S. Thorne and J. A. Wheeler, {\it Gravitation} (W. H. Freeman, San Francisco, 1973)

\bibitem{chernoff1983mixmaster}
D. F. Chernoff and J. D. Barrow, "Chaos in the Mixmaster Universe", {\it Phys. Rev. Lett.} {\bf 50} (1983) 134

\bibitem{cornish1997mixmaster}
N. J. Cornish and J. J. Levin, "Mixmaster universe: A chaotic Farey tale", {\it Phys. Rev.} D {\bf 55} (1997) 7489\\
N. J. Cornish and J. J. Levin, "The mixmaster universe is chaotic", {\it Phys. Rev. Lett.} {\bf 78} (1997) 998\\
M. Bojowald, D. Brizuela, P. C. Cabrera and S. F. Uria, "Chaotic behavior of the Bianchi IX model under the influence of quantum effects", {\it Phys. Rev.} D {\bf 109} (2024) 044038

\bibitem{corichi2007polymer}
A. Corichi, T. Vukasinac and J. A. Zapata, ``Polymer quantum mechanics and its continuum limit'', {\it Phys. Rev.} D {\bf 76} (2007) 044016 (arXiv: 0704.0007 [gr-qc])

\bibitem{ashtekar2003mathematical}
A. Ashtekar, S. Fairhurst and J. L. Willis, "Quantum gravity, shadow states, and quantum mechanics", {\it Class. Quant. Grav.} {\bf 20} (2003) 1031 (arXiv: gr-qc/0207106)

\bibitem{CPR}A. Corichi and T. Vuka\v{s}inac, "Effective constrained polymeric theories and their continuum limit", {\it Phys. Rev.} D {\bf 86} (2012) 064019 (arXiv: 1202.1846 [gr-qc])

\bibitem{banerjee2007discreteness}
K. Banerjee and G. Date, "Loop Quantization of Polarized Gowdy Model on $T^3$: Classical Theory", {\it Class. Quant. Grav.} {\bf 25} (2008) 105014 (arXiv: 0712.0683 [gr-qc])

\bibitem{NaturalCutoff} K. Nozari, M. A. Gorji, V. Hosseinzadeh and B. Vakili, "Natural cutoffs via compact symplectic manifolds", {\it Class. Quantum Grav.} {\bf 33} (2016) 025009 (arXiv:
1405.4083 [gr-qc])

\bibitem{bojowald2001absence}
M. Bojowald, "Absence of singularity in loop quantum cosmology", {\it Phys. Rev. Lett.} {\bf 86} (2001) 5227 (arXiv: gr-qc/0102069)

\bibitem{qc pol}G. De Risi, R. Maartens and P. Singh, "Graceful exit via polymerization of pre-big bang cosmology", {\it Phys. Rev.} D {\bf 76} (2007) 103531 (arXiv: 0706.3586
[hep-th])\\S. M. Hassan and V. Husain, "Semiclassical cosmology with polymer matter", {\it  Class. Quantum Grav.} {\bf 34} (2017) 084003 (arXiv: 1705.00398 [gr-qc])\\B. Vakili, K. Nozari, V. Hosseinzadeh and M. A. Gorji, "Bouncing scalar field cosmology in the polymeric minisuperspace picture", {\it  Mod. Phys. Lett.} A {\bf 29} (2014) 1450169
(arXiv: 1408.4535 [gr-qc])\\J. B. Achour, F. Lamy, H. Liu and K.
Noui, "Polymer Schwarzschild black hole: An effective metric", {\it Europhys. Lett.} {\bf 123} (2018) 20006 (arXiv:
1803.01152 [gr-qc])\\ L. Boldorini, C. Marzano and G. Montani, "Polymer Geodesic Motion in Schwarzschild Spacetime", (arXiv: 2506.07732 [gr-qc])\\B. Vakili, "Classical polymerization of the Schwarzschild metric", {\it Adv. High Energy Phys.} (2018) 3610543 (arXiv: 1806.01837 [hep-th])

\bibitem{pol bl}A. Peltola and G. Kunstatter, "Effective Polymer Dynamics of D-Dimensional Black Hole Interiors", \emph{Phys. Rev.} D {\bf 80} (2009) 044031 (arXiv:
0902.1746 [gr-qc])\\E. Bianchi, "Black Hole Entropy, Loop Gravity, and Polymer Physics", \emph{Class. Quantum Grav.}
{\bf 28} (2011) 114006 (arXiv: 1011.5628 [gr-qc])\\M. A. Gorji, K. Nozari and B. Vakili, "Polymeric Quantization and Black Hole Thermodynamics", {\it  Phys.
Lett.} B {\bf 735} (2014) 62 (arXiv: 1312.6835 [gr-qc])\\M. A. Gorji,
K. Nozari and B. Vakili, "Thermostatistics of the Polymeric Ideal Gas", {\it Phys. Rev.} D {\bf 90} (2014) 044051
(arXiv: 1408.1725 [gr-qc])\\L. Boldorini and G. Montani, "Effective Quantum Gravitational Collapse in a Polymer Framework", {\it J. Cosmol. Astropart. Phys.} {\bf 10} (2024) 090 (arXiv: 2406.03279 [gr-qc])\\
 M. V. Battisti, O. Maria Lecian and G. Montani, "Polymer Quantum Dynamics of the Taub Universe", {\it Phys. Rev.} D {\bf 78} (2008) 103514 (arXiv: 0806.0768 [gr-qc])


\bibitem{GUP Noz} K. Nozari and A. Etemadi, "Minimal length, maximal momentum and Hilbert space representation of quantum mechanics", {\it Phys. Rev.} D {\bf 85} (2012) 104029
(arXiv: 1205.0158 [hep-th])\\ A. N. Tawfik and A. M. Diab, "A review of the generalized uncertainty principle", {\it Rep. Prog. Phys.} {\bf 78} (2015) 126001

\bibitem{snyder}H. S. Snyder, "Quantized space-time", {\it Phys. Rev.} {\bf 71} (1947) 38\\
S. Meljanac, D. Meljanac, A. Samsarov and M. Stojic, "Kappa Snyder deformations of Minkowski spacetime, realizations and Hopf algebra", {\it Phys. Rev.} D {\bf 83} (2011) 065009 (arXiv: 1102.1655 [math-ph])\\M. A. Gorji, K. Nozari and B. Vakili, "Polymer quantization versus the Snyder noncommutative space", {\it Class. Quantum Grav.} {\bf 32} (2015) 155007 (arXiv: 1506.03375 [gr-qc])

\bibitem{kon} M. Kontsevich, "Deformation Quantization of Poisson Manifolds", {\it Lett. Math. Phys.} {\bf 66} (2003) 157 (arXiv: q-alg/9709040)\\T. Nakamura, 
{\it The deformations of symplectic structures by moment maps}, (arXiv: 1605.02448 [math.DG])

\bibitem{vakilibianki} G. Barca and S. Gielen, {\it Bouncing Bianchi Models with Deformed Commutation Relations} (arXiv: 2507.01678 [gr-qc])\\
B. Vakili, "Classical Polymerization of the Bianchi I Model with Deformed Poisson Structure", {\it Int. J. Geom. Methods Mod. Phys.}, in press, https://doi.org/10.1142/S0219887826500349, (arXiv: 2510.06628 [gr-qc])
\end{thebibliography}
\end{document}